\documentclass[journal,12pt,draftclsnofoot,onecolumn]{IEEEtran}
\IEEEoverridecommandlockouts
\ifCLASSINFOpdf
\else
\fi
\usepackage[cmex10]{amsmath}
\usepackage{amsthm,cite,graphicx,,booktabs,lipsum,color,bm,algorithm,subcaption,soul}
\usepackage{algpseudocode,pifont,tikz,paralist,multirow,xcolor}

\begin{document}
\begin{titlepage}
\begin{center}
\vspace*{-2\baselineskip}
\begin{minipage}[l]{7cm}
\flushleft
\includegraphics[width=2 in]{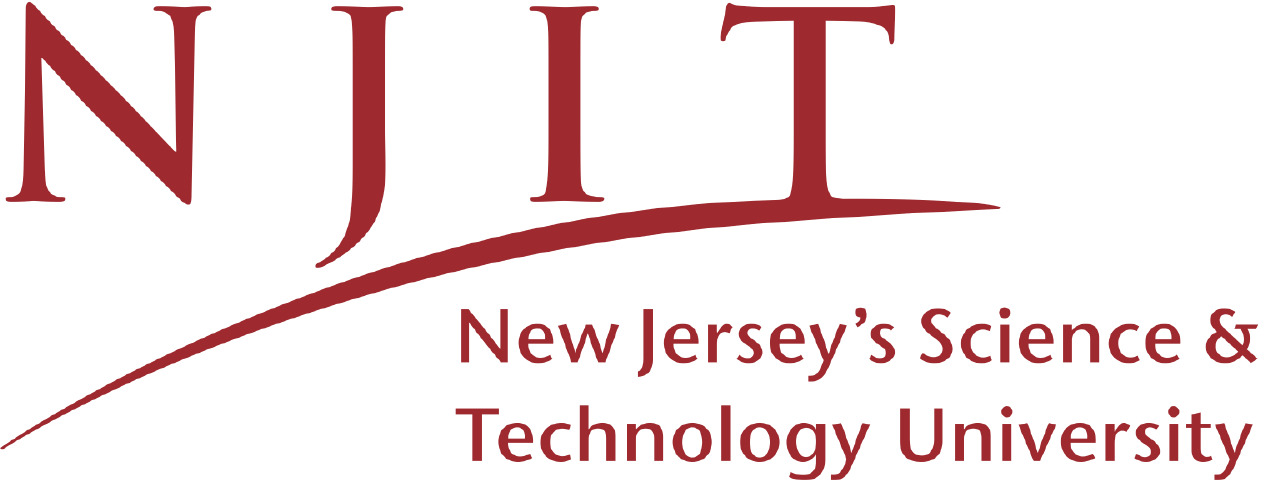}
\end{minipage}
\hfill
\begin{minipage}[r]{7cm}
\flushright
\includegraphics[width=1 in]{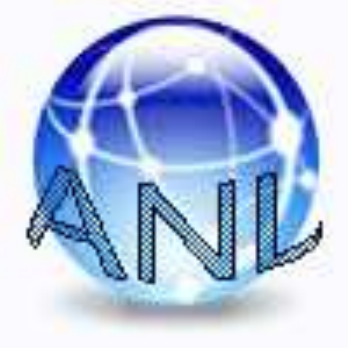}
\end{minipage}

\vfill

\textsc{\LARGE Mobile Edge Computing Empowers Internet of Things\\[12pt]}
\vfill
\textsc{%\LARGE Author Name\\[12pt]
\LARGE  NIRWAN ANSARI\\XIANG SUN}\\
\vfill
\textsc{\LARGE TR-ANL-2017-003\\[12pt]
\LARGE May 11, 2017}\\[1.5cm]
% Bottom of the page
\vfill
{ADVANCED NETWORKING LABORATORY\\
 DEPARTMENT OF ELECTRICAL AND COMPUTER ENGINEERING\\
 NEW JERSY INSTITUTE OF TECHNOLOGY}
\end{center}
\end{titlepage}
\title{Mobile Edge Computing Empowers Internet of Things}

\author{Nirwan~Ansari,~\IEEEmembership{Fellow,~IEEE,}
       Xiang~Sun,~\IEEEmembership{Student~Member,~IEEE}  
\thanks{N. Ansari and X. Sun are with Advanced Networking Lab., Department of Electrical $\&$ Computer Engineering, New Jersey Institute of Technology, Newark, NJ 07102, USA. E-mail:$\{$nirwan.ansari, xs47$\}$@njit.edu.\newline
Corresponding author: Xiang Sun.\newline
The paper has been accepted by IEICE Transactions on Communications.\newline}% <-this % stops a space <-this % stops a space
%\thanks{Manuscript received April 19, 2005; revised September 17, 2014.}
}

%\markboth{Journal of \LaTeX\ Class Files,~Vol.~13, No.~9, September~2014}%
%{Shell \MakeLowercase{\textit{et al.}}: Bare Demo of IEEEtran.cls for Journals}

\maketitle

\begin{abstract}
In this paper, we propose a Mobile Edge Internet of Things (MEIoT) architecture by leveraging the fiber-wireless access technology, the cloudlet concept, and the software defined networking framework. The MEIoT architecture brings computing and storage resources close to Internet of Things (IoT) devices in order to speed up IoT data sharing and analytics. Specifically, the IoT devices (belonging to the same user) are associated to a specific proxy Virtual Machine (VM) in the nearby cloudlet. The proxy VM stores and analyzes the IoT data (generated by its IoT devices) in real-time. Moreover, we introduce the semantic and social IoT technology in the context of MEIoT to solve the interoperability and inefficient access control problem in the IoT system. In addition, we propose two dynamic proxy VM migration methods to minimize the end-to-end delay between proxy VMs and their IoT devices and to minimize the total on-grid energy consumption of the cloudlets, respectively. Performance of the proposed methods is validated via extensive simulations. 
\end{abstract}


\section{Introduction}\label{sec:Introduction}
Internet of Things (IoTs) is enabling interconnections among a tremendous number of things such that different things can share their observations of the physical world. According to a new Gartner forecast, 26 billion things (excluding PCs, tablets, and smartphones) will be installed in 2020 \cite{Gartner}. Cisco predicted that 50 billion devices will be connected to the Internet by 2020 \cite{Evans:2011:IOT}. These connected things will generate a humongous volume of data, which digitally represent the states of the physical world. However, owing to the resource constrained nature, many IoT devices cannot always guarantee the interconnections, i.e., some IoT devices cannot be reachable owing to their periodical sleep schemes and intermittent wireless connections. Thus, it is important to design an efficient mechanism to facilitate resource constrained IoT devices in sharing their data over the network. Also, the resource constrained IoT devices cannot feasibly conduct complicated data access management, thus sharing IoT data over the network poses serious security challenges, i.e., unauthorized users/devices may easily access and misuse the shared IoT data, which may contain personal information. Therefore, designing an efficient access control mechanism tailored for IoT devices is critical to empowering the current IoT system. In addition, only providing interconnections to share raw data among IoT devices is not enough to gain the insight behind the big IoT data; the insight is more valuable for the society as a whole. Thus, it is beneficial to provision the IoT system with a comprehensive cognitive capability such that high-level knowledge can be extracted from the big IoT raw data streams by applying various types of data mining and machine learning methods \cite{Wu:2014:CIT}. The data center infrastructure has been demonstrated to provision resources flexibly and efficiently \cite{Sun:2016:ORU}; meanwhile, various parallel computing architectures and distributed storage frameworks have been designed based on a data center (e.g., MapReduce \cite{MapReduce} and Spark \cite{Spark}). Thus, it is desirable to transmit the big IoT data from the IoT devices to remote data centers via the Internet for further data analysis. Yet, this would place a heavy burden on the network to conduct data aggregation from IoT devices to a centralized data center, and thus exponentially increase the network delay for transmitting big IoT data to remote data centers, especially during the peak time. Note that delay is a key performance metric in provisioning the Quality of Service (QoS) for many IoT applications. For instance, smart grid applications have stringent requirement on latency up to 20 ms; processing automation applications (i.e., monitoring and diagnosing of industrial elements and processes) imposes latency requirements ranging from 50 ms to 100 ms \cite{Schulz:2017:LCI}.

The Mobile Edge Computing (MEC) concept \cite{Mach:2017:MEC} is essentially bringing the computing and storage capability from remote data centers to the mobile edge in order to reduce the network delay between end devices and computing resources. Specifically, various computing resources, attached to edge routers, wireless access points (WAPs), and smart gateways, are available for nearby mobile devices; thus, these devices can offload their workloads to computing resources at the edge, thus potentially reducing the energy consumption of the devices and accelerating the computing processes \cite{Sun:205:EBA}. Empowering IoT with MEC can essentially improve the QoS for IoT applications. Basically, IoT applications, which try to obtain the corresponding data from different types of IoT devices and generate high-level knowledge by analyzing the acquired data based on data analytic models, would be deployed at the mobile edge, and thus the data streams generated by the IoT devices would be uploaded to the IoT applications without traversing the mobile core network. This can significantly alleviate the traffic load in the core network and potentially speed up the IoT applications in processing big IoT data streams.

In this paper, we will design a novel Mobile Edge IoT (MEIoT) architecture by leveraging the Fiber-Wireless (FiWi) access technology, the mobile network, the cloudlet concept, and the Software Defined Networking (SDN) framework to efficiently share and analyze the big IoT data at the mobile edge. The rest of the paper is organized as follows. In Sec. II, we describe the proposed MEIoT architecture. In Sec. III, we illustrate the four challenges of the current IoT system and propose four potential solutions tailored for the MEIoT architecture. In Sec. IV, we evaluate the performance of the proposed solutions via simulations. We briefly delineate the future work in Sec. V, and present the conclusion in Sec. VI.   

\begin{figure*}[!htb]
\begin{minipage}[t]{1\linewidth}
	\centering	
	\includegraphics[width=1.0\columnwidth]{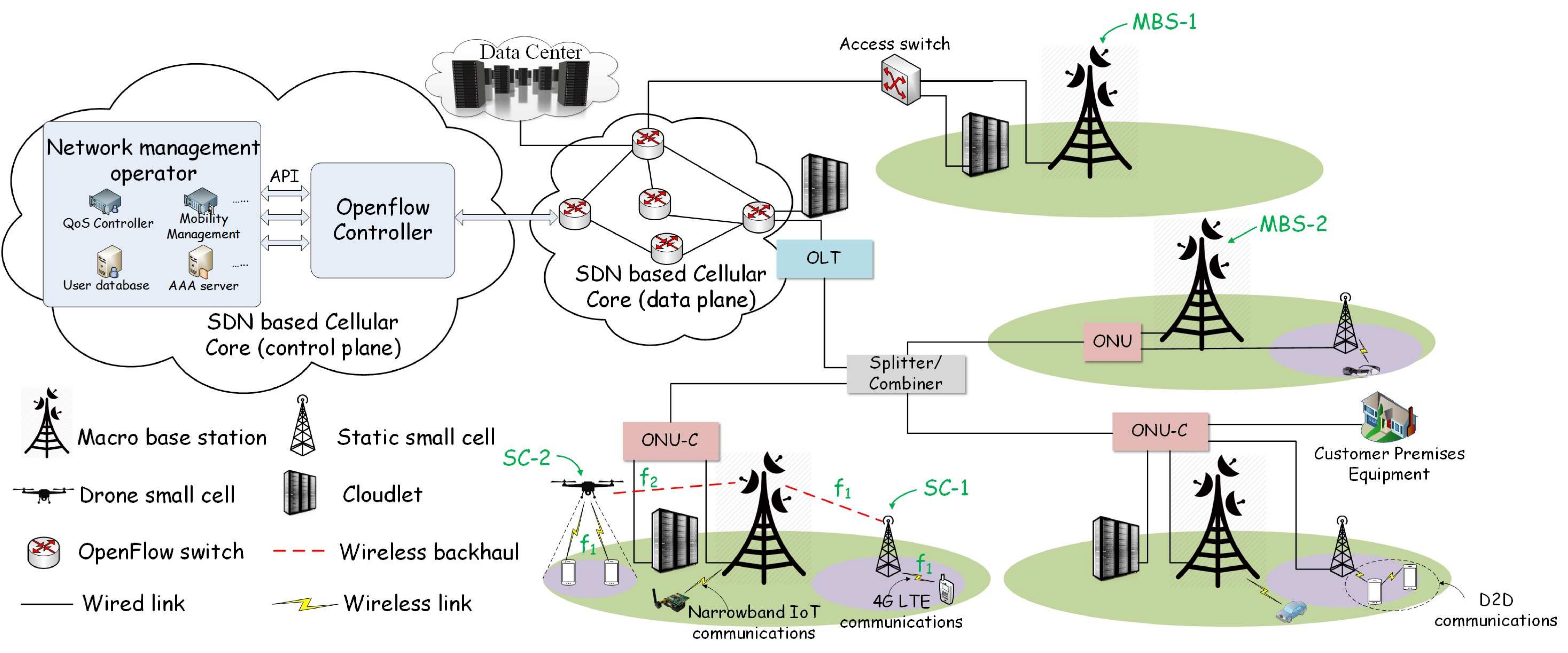}
	\caption{The MEIoT architecture.}	
	\label{fig:overall_MEIoT}
\end{minipage}	
\end{figure*}

\section{MEIoT architecture}
In order to facilitate IoT data sharing and analytics, we propose the MEIoT architecture, as shown in Fig. \ref{fig:overall_MEIoT}. The MEIoT architecture comprises five parts, i.e., multi-interface wireless access network, heterogeneous backhauling, distributed cloudlets, hierarchical structure of a cloudlet, and the SDN based mobile core network. We will next detail these five parts.

\subsection{Multi-interface wireless access network}
Various WAPs, such as WiFi access points, static Small Cells (SC) (e.g., pico cell, femto cell, etc.), dynamic SCs (e.g., drone mounted SCs), and Macro BSs (MBSs), have already been deployed in the mobile network and provide high radio coverage and network capacity. Thus, distributed WAPs have the potential to connect all IoT devices whether they are moving or static. Yet, different IoT devices have different communications requirements; that is, some energy-sensitive IoT devices (e.g., smart meters) require very low transmission date rate and some energy-insensitive devices (e.g., surveillance devices and mobile phones) need high-speed transmission to meet their embedded application requirements. The heterogeneous data transmission requirements among IoT devices effectuate different devices to adopt different wireless technologies (e.g., D2D communications, NarrowBand IoT communications, LTE communications, etc.) to share their sensed data. Thus, WAPs are equipped with multiple wireless access interfaces such that they can communicate with IoT devices by applying different wireless technologies. 

%Different wireless access points may use different backhaul technologies to transmit/receive data to/from the mobile core network. For example, drone-based SCs often apply wireless backhaul technologies (e.g., millimeter-wave, microwave, and sub-6 GHz unlicensed/licensed) to communicate with corresponding Macro BSs (MBSs), and WiFi access points and MBSs can use optical backhaul (e.g., Ethernet passive optical network) technologies to communicate with mobile core network.

\begin{figure*}[!htb]
\begin{minipage}[t]{1\linewidth}
	\centering	
	\includegraphics[width=1.0\columnwidth]{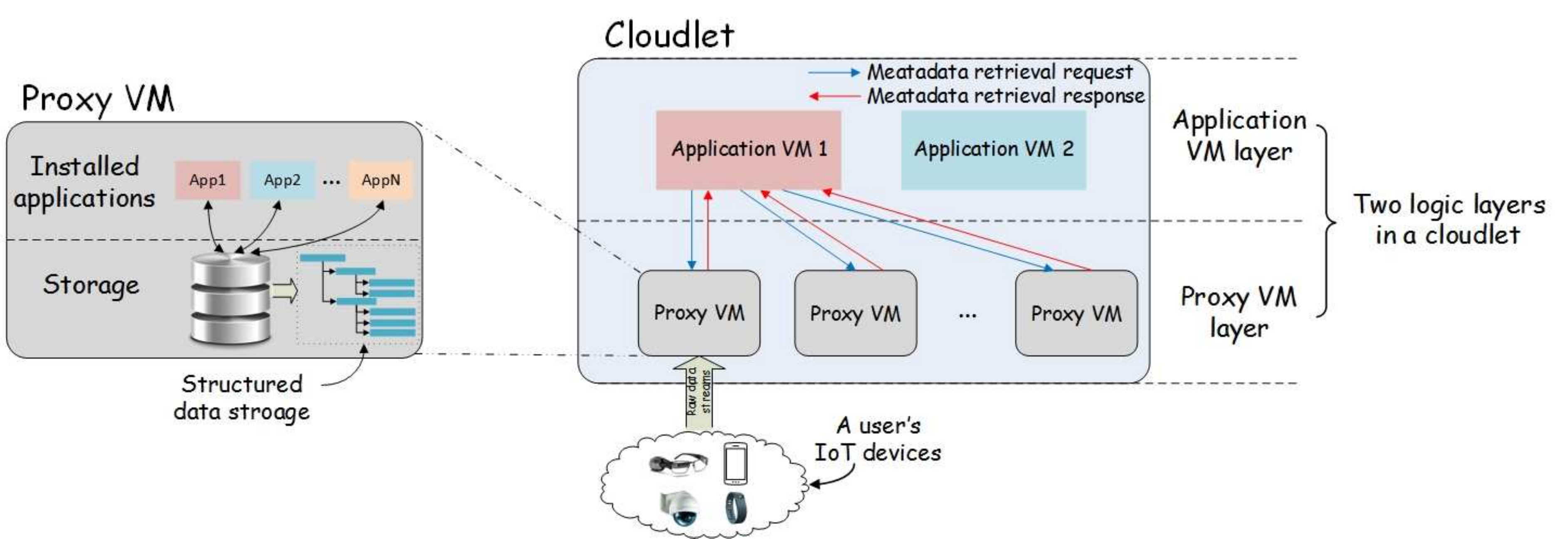}
	\caption{The hierarchical structure within a cloudlet.}	
	\label{fig:cloudlet_framework}
\end{minipage}	
\end{figure*}

\subsection{Heterogeneous backhauling} 

The mobile backhaul is used to carry the traffic from WAPs to the mobile core network \cite{Clarke:2014:EMW}. Owing to different requirements (such as low latency, long distance transmission, mobility, and high reliability), various backhaul technologies have been proposed. The current backhaul technologies can mainly be divided into two categories, i.e., wired and wireless backhaul solutions.

\subsubsection{Wired backhaul solutions}

Wired backhaul provides the advantages of high reliability, high data rate, and high availability. A WAP (such as MBS-1 in Fig. \ref{fig:overall_MEIoT}) can communicate with the mobile core network based on a wired connection (e.g., xDSL), which traverses an access switch. The access switch is connected to a cloudlet and conducts L2/L3 switching among the WAP, the cloudlet, and the mobile core network. A cloudlet\cite{Satyanarayanan:2009:TCV}, which comprises a number of interconnected Physical Machines (PMs), provides computing and storage resources to IoT devices with low latency.

Passive optical networks (PONs) can potentially provision cloud computing \cite{Taheri:2013:FSP}, and thus a WAP can also utilize the optical backhaul to achieve extra low communications delay. For example, MBS-2 in Fig. \ref{fig:overall_MEIoT} is connected to an Optical Network Unit (ONU), which is further connected to an Optical Line Terminal (OLT) via an optical splitter/combiner. The function of an ONU is to aggregate the traffic from its connected WAPs and communicate with its connected OLT based on the assigned wavelength channels. The function of an OLT is to provide L2/L3 switching between mobile core network and its connected ONUs. Note that there are two different types of ONUs in the MEIoT architecture, traditional ONU and ONU-Cloudlet (ONU-C). Different from traditional ONUs, an ONU-C, which is normally connected to a local cloudlet, can not only relay the traffic between its connected OLT and WAPs but also provide the switching function to enable the local communications between its WAPs and its connected cloudlet \cite{Rimal:2017:MEC}. Thus, the communications between the local WAPs and the cloudlet can be offloaded from the OLT and the mobile core network. This can significantly reduce the traffic load of the mobile core network and the OLT.  

\subsubsection{Wireless backhaul solutions}
Wireless backhual technologies present the advantages of flexible deployment and low cost. Currently, many SCs (especially for mobile SCs, such as drone mounted SCs \cite{Zhang:2017:SSD}) apply the wireless backhaul solutions to facilitate the communications between MBSs and SCs. The wireless backhaul solutions can be divided into two categories: in-band and out-band wireless backhaul.    

In-band wireless backhaul means that the wireless bands being applied to the communications between an SC and its mobile users are the same as those being applied to the wireless backhaul between the SC and its MBS \cite{Li:2015:SCI}. For instance, SC-1 in Fig. \ref{fig:overall_MEIoT} uses the same band of $f_1$ to communicate with both the MBS and the mobile users. The in-band wireless backhaul achieves high frequency utilization, but it requires efficient scheduling to reduce the interference between the wireless backhaul channels and the downlink/uplink channels between an SC and its mobile users.

Out-band wireless backhaul implies that the wireless backhaul of an SC applies bands different from those used for the communications between the SC and its mobile users. For instance, SC-2 in Fig. \ref{fig:overall_MEIoT} applies $f_2$ band to communicate with the MBS and uses $f_1$ band to communicate with its mobile users. Currently, two technologies have been proposed for the out-band wireless backhaul, i.e., millimeter wave (mmWave)\cite{Hur:2013:MWB} and sub-6 GHz wireless backhaul. mmWave offers high capacity and reliability based on line-of-sight communications between an MBS and an SC, and sub-6 GHz wireless backhaul provides high data rate based on non-line-of-sight communications between an MBS and an SC \cite{Zhang:2016:FHB}. 

\subsection{Distributed cloudlets} 

Cloudlets are deployed at the mobile edge to provision computing and storage resources to IoT devices with low latency. The data generated by various IoT devices can be stored and analyzed in the corresponding cloudlets in real time. The deployment of cloudlets is flexible, i.e., a cloudlet can connect to an ONU-C/access switch such that the IoT devices, which are associated with the WAPs (which are directly connected to the ONU-C/access switch), can utilize the computing and storage resources of the cloudlet without traversing OLTs and the mobile core network. Also, a cloudlet can be deployed at the edge of the mobile core network or connected to an OLT such that more WAPs can share the computing and storage resources in the same cloudlet. Note that geographical distributed cloudlets can share their resources with each other via the mobile core network.

Remote data centers are located at remote sites (which are mostly directly connected to the core network) to provide the scalability and availability of the system. Specifically, the computing and storage capacities of the local cloudlets are limited, and thus they may not have enough capacities to efficiently store and analyze IoT data streams. Data centers, which supply sufficient and flexible resource provisioning, can be considered as backup units to store and analyze IoT data streams.

It is possible that either mobile network operators or cloud vendors are willing to deploy cloudlets in the network to facilitate their business. For instance, Nokia has established a Multi-access Computing platform to attach a cloudlet to an LTE MBS. Based on the platform, a use case named ``connected cars'' has been developed, where each local cloudlet analyzes the data at the point of capture and feeds back the insights to the vehicles within the cloudlet's coverage with extremely low latency (less than 20 $ms$) in order to improve road safety \cite{MEC:NOKIA}. Also, Microsoft has shown great interest in deploying cloudlets at the network edge to reduce the latency of mobile devices in accessing computing resources and improve the battery life of mobile devices. It has also suggested to build an extensive infrastructure of micro data centers (i.e., cloudlets) including 1-10 servers with several terabytes of storage (which may cost \$20K-\$200K per cloudlet) and place them everywhere \cite{EMD:2015:Bahl}. Building a large number of cloudlets at the mobile edge may incur a huge capital expenditure; however, it would potentially generate huge revenue by renting out local computing and storage resources to users and application providers.            

Different cloudlet providers may own different cloudlets, and sharing computing and storage resources among cloudlets (which are owned by different providers) can facilitate the resource provisioning to IoT devices. Thus, a fair pricing model needs to be established to achieve resource sharing among different cloudlet providers.     	

\subsection{Hierarchical structure of a cloudlet} 

As shown in Fig. \ref{fig:cloudlet_framework}, there are two logical layers in a cloudlet, i.e., Proxy VM layer and Application VM layer. Proxy VM layer comprises a number of the proxy VMs. A proxy VM is considered as a private VM associated with the IoT devices, which belong to the same user\footnote{A user can be a person who owns various private IoT devices, an entity/company that deploys a set of IoT devices in the area (such as the surveillance cameras), or a group of users who trust each other and share the same proxy VM.}. Specifically, IoT devices of a user would be registered to the user's proxy VM. After the registration, these IoT devices would upload their sensed data to the proxy VM periodically or upon requests. The proxy VM converts these raw data into structured data, stores them in the local storage space, conducts the access control policies to protect stored data, and/or pre-processes the structured data upon requests. The application VM layer comprises a number of application VMs, which are deployed by the application providers, to retrieve metadata from proxy VMs, analyze the received data to generate high-level knowledge, and provide the corresponding service to users. 

Here, we provide the terrorist detection application to illustrate how the hierarchical cloudlet architecture works. First, if some users are interested in the service provided by the terrorist detection application, they can install the corresponding app (e.g., App1 in Fig. \ref{fig:cloudlet_framework}) in their proxy VMs. Local devices upload their captured photos/videos to their proxy VMs over time. If the terrorist detecting application VM tries to locate a specific terrorist by conducting face matching over the captured photos/videos, instead of having each proxy VM transmit its captured photos/videos to the application VM, the terrorist detecting application VM would send a metadata retrieval request containing the terrorist's photo to the proxy VMs (which have installed the corresponding app) among all the cloudlets. After receiving the request, the installed app in the proxy VM would retrieve the videos/photos in the local storage, and conduct the face matching algorithm by comparing these videos/photos with the received terrorist's photo. If a match is detected, the proxy VM would respond to the application VM with the related metadata, i.e., the location information and time stamps of the corresponding photos/videos.

The hierarchical cloudlet architecture can also facilitate other applications, such as ParkNet \cite{Mathur:2010:ParkNet} and FaceDate \cite{Neog:2016:FaceDate}.
\begin{itemize}[leftmargin=*]
\item{\ul{ParkNet} helps users locate available parking spots in the urban area. Specifically, the proxy VMs collect the sensed data streams from their smart cars, which are considered as IoT devices and are registered to the corresponding proxy VMs. Note that each smart car is equipped with a GPS receiver and a passenger-side-facing ultrasonic rangefinder to generate the location and parking spot occupancy information. Each proxy VM analyzes the information and generates the metadata, which identify the available parking spots, and forwards the metadata to the application VM. The application VM will inform and assign the available parking spots to the smart cars upon requests.}
\item{\ul{FaceDate} is to find and date nearby people based on their face preference in real-time. Specifically, each user uploads a profile photo and provides basic information (such as date of birth, gender, and a brief write-up) about himself/herself into its proxy VM. In addition, each user uploads a set of preference photos (i.e., the photos of a boy/girl whom she/he wants to date. For instance, if a man wants to date a woman who resembles Marilyn Monroe, he would upload the photos of Marilyn Monroe into its proxy VM) to identify his dream date partner. If a man tries to find a nearby date partner, his proxy VM (i.e., request proxy VM) would send a request, which contains its preference photos, to the application VM. The application VM forwards the request to other proxy VMs, which conduct the face recognition algorithm by comparing the preference photos in the request with their users' profile photos. If the photos are highly matched, the proxy VMs (i.e., response proxy VMs) would respond to the application VM with the metadata (similarity of the photos) as well as the preference photos in these response proxy VMs. The application VM would forward these preference photos (from the response proxy VMs) to the request proxy VM, which conducts the face recognition algorithm by comparing the received preference photos with its profile photo and responds to the application VM with the metadata (similarity of the photos). Finally, the application VM would pick the best matched candidate and enable the chatting accordingly.}                     
\end{itemize}

The proposed hierarchical cloudlet architecture exhibits the following advantages:
\begin{itemize}[leftmargin=*]
\item{\emph{Simplified IoT devices}: Each IoT device only needs to sense the environment and upload the sensed data to its proxy VM, which converts the raw data into structured data, stores them into local storage, shares the local structured data by responding to data retrieval requests from other devices, conducts the access control, and pre-processes the structured data by converting them into metadata. Thus, associating IoT devices to their proxy VM is essentially moving most of the functionalities, which are originally executed at the IoT devices, to their proxy VM. This can significantly reduce the energy consumption of IoT devices and speed up the IoT data sharing/analytics process.}
\item{\emph{Accessibility of IoT data}: Cloudlets are connected with each other via wired links, implying that IoT data stored in the cloudlets are always accessible. This resolves the weak accessibility problem in the traditional IoT system, where IoT data are stored in the IoT devices and application VMs cannot retrieve IoT data from IoT devices which may periodically sleep and suffer from intermittent wireless connections.}
\item{\emph{Privacy preserving}: IoT data, which are generated by IoT devices of a user, are basically stored in the user's proxy VM, which is considered as a private VM to facilitate resource isolation and access control. Moreover, the proxy VM can pre-process the IoT data to share metadata (rather than raw data) by removing the user's personal information from raw data. For instance, in the terrorist detection application, each proxy VM only provides the locations and the time stamps of the matched photos/videos rather than the photos/videos.}
\item{\emph{Efficient distributed computing}: Each proxy VM becomes a worker node of an application VM, which acts as a master node to distribute workloads to proxy VMs, aggregate metadata from proxy VMs, and provide services to users. This distributed computing structure can fully utilize the distributed computing resources in the proxy VMs and significantly reduce the traffic load of the network as compared to the current way in which the application VM retrieves the raw data from IoT devices, analyzes them and provides services to users.}
\end{itemize}

\subsection{SDN based mobile core network}

Instead of applying the traditional cellular core network architecture, which leads to inefficient, inflexible, and unscalable packet forwarding, the SDN based mobile core network \cite{Sun:2016:EdgeIoT,Jin:2013:Softcell,Sun:2016:PRIMAL,Sun:2015:GEA,Sun:2017:GCN,Sun:2017:LAW,Sun:2017:AAH} is adopted in MEIoT. The SDN based mobile core network is essentially decoupling the control plane from the switches, which only run data plane functionalities. The control plane is offloaded to a logical central controller, which transmits the control information (e.g., flow tables) to the OpenFlow switches by applying the OpenFlow protocol \cite{Openflow}, monitors the traffic statistics of the network, and provides Application Programming Interfaces (APIs) to network management operators so that different mobile network functionalities, such as mobility management, user authentication, authorization and accounting, network virtualization, and QoS control, can be added, removed, and modified flexibly.

\section{Challenges and solutions in MEIoT}
In this section, we will discuss some challenges in realizing real-time IoT data sharing/analytics and provide some potential solutions in the context of MEIoT.
\subsection{Challenge 1: Interoperability problem during IoT data sharing}
\begin{figure}[t]
	\centering	
	\includegraphics[width=0.9\columnwidth]{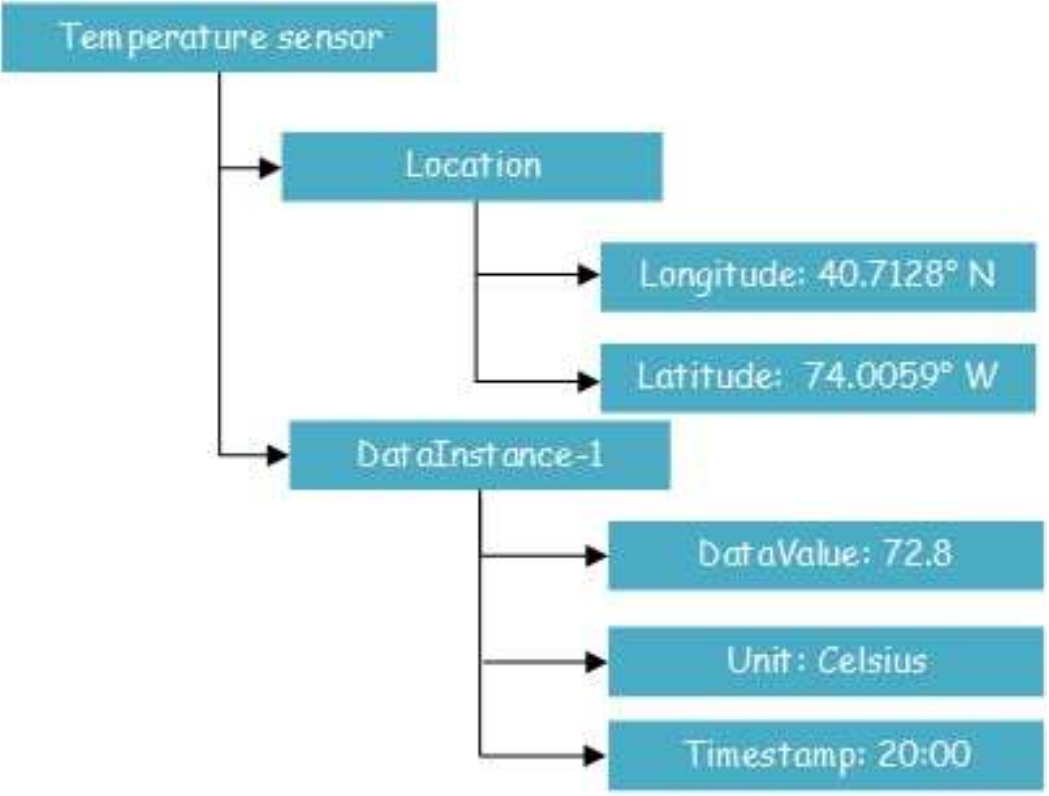}
	\caption{One example of structured data.}	
	\label{fig:structured_data}	
\end{figure}

Sharing IoT data among devices is the basic objective in the IoT system. Traditionally, if a client (i.e., an IoT device requests data from another IoT device) tries to obtain the data from a server (i.e., an IoT device generates IoT data), the client would send a request to the server, which would respond to the client with the corresponding data. In the MEIoT architecture, proxy VMs are considered as the gateways to store and manage the IoT data streams from their registered IoT devices. Normally, all the clients would send the requests to the corresponding proxy VMs (rather than the original devices, which generate IoT data) in order to retrieve their IoT data. Proxy VMs receive the raw data streams from their registered IoT devices, convert them into structured data, and respond to the data retrieval requests. Fig. \ref{fig:structured_data} just provides one example to illustrate the structured data provided by a temperature sensor. 

However, sharing the structured data among devices provides weak interoperability because different proxy VMs and IoT devices may apply different data models and vocabularies to annotate and structurize IoT data. For example, as shown in Fig. \ref{fig:structured_data}, the proxy VM applies ``DataValue'' to annotate a temperature value\footnote{Normally, a proxy VM would use the same vocabularies that are applied by its registered devices. For example, if the temperature sensor uses ``DataValue'' to annotate its captured temperature value, the proxy VM would apply the same vocabulary to annotate the temperature value.}. However, a client (who tries to retrieve the current temperature value) uses ``TempValue'' to annotate a temperature value, and thus the client would request to obtain the ``TempValue'' of the temperature sensor. Obviously, the proxy VM cannot understand the request from the client. This interoperability problem hampers the IoT data from being shared among different devices.

\subsection{Solution 1: Semantic interoperability in MEIoT}
\subsubsection{Basic concept of semantics}
\begin{figure*}[!htb]
\centering
\begin{subfigure}{.4\textwidth}
	\centering	
	\includegraphics[width=0.9\columnwidth]{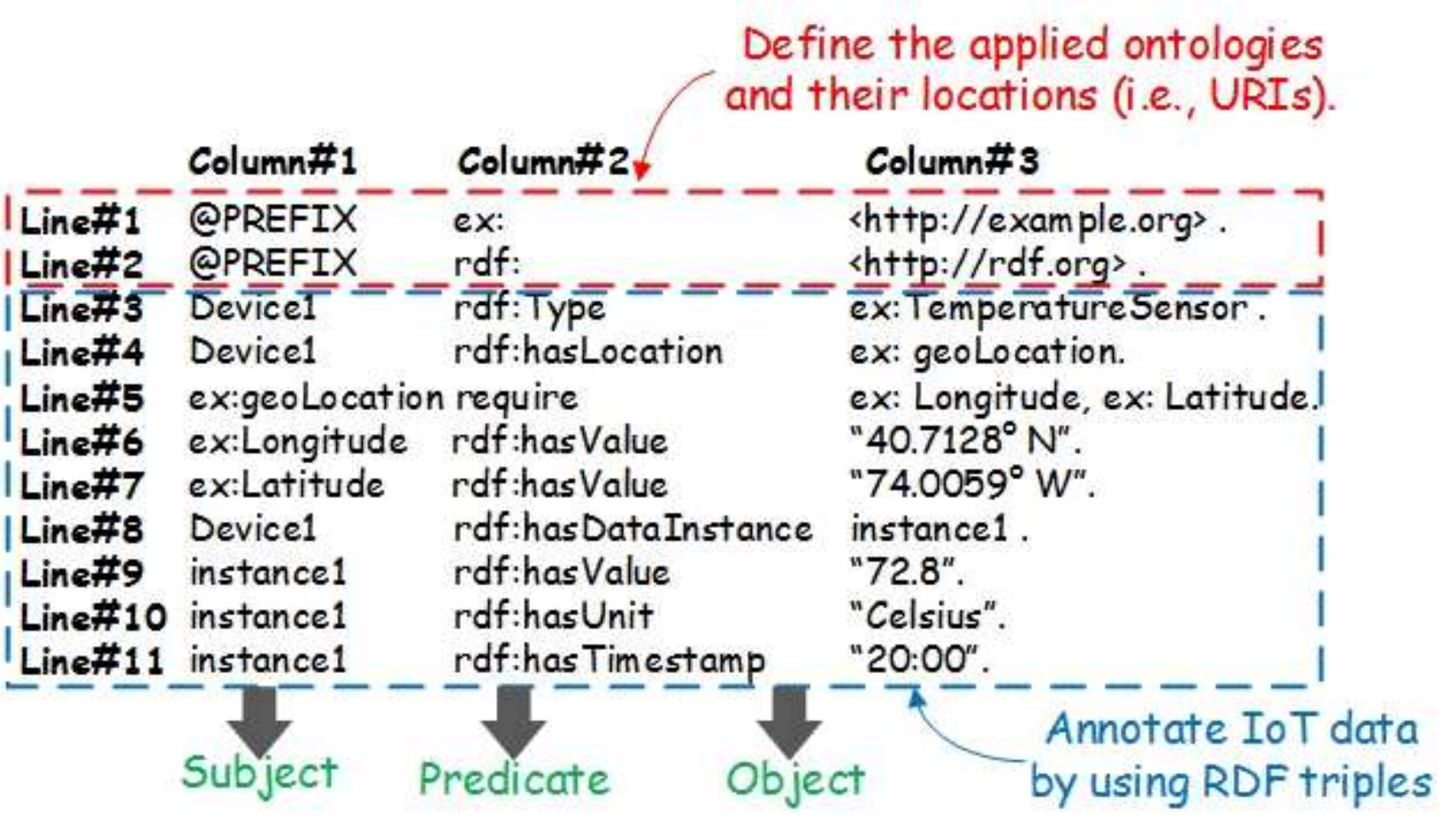}
	\caption{One example of semantic data represented by RDF triples.}	
	\label{fig:semantic_data}	
\end{subfigure}%
\hspace{0.1cm}
\begin{subfigure}{.5\textwidth}
	\centering	
	\includegraphics[width=1.05\columnwidth]{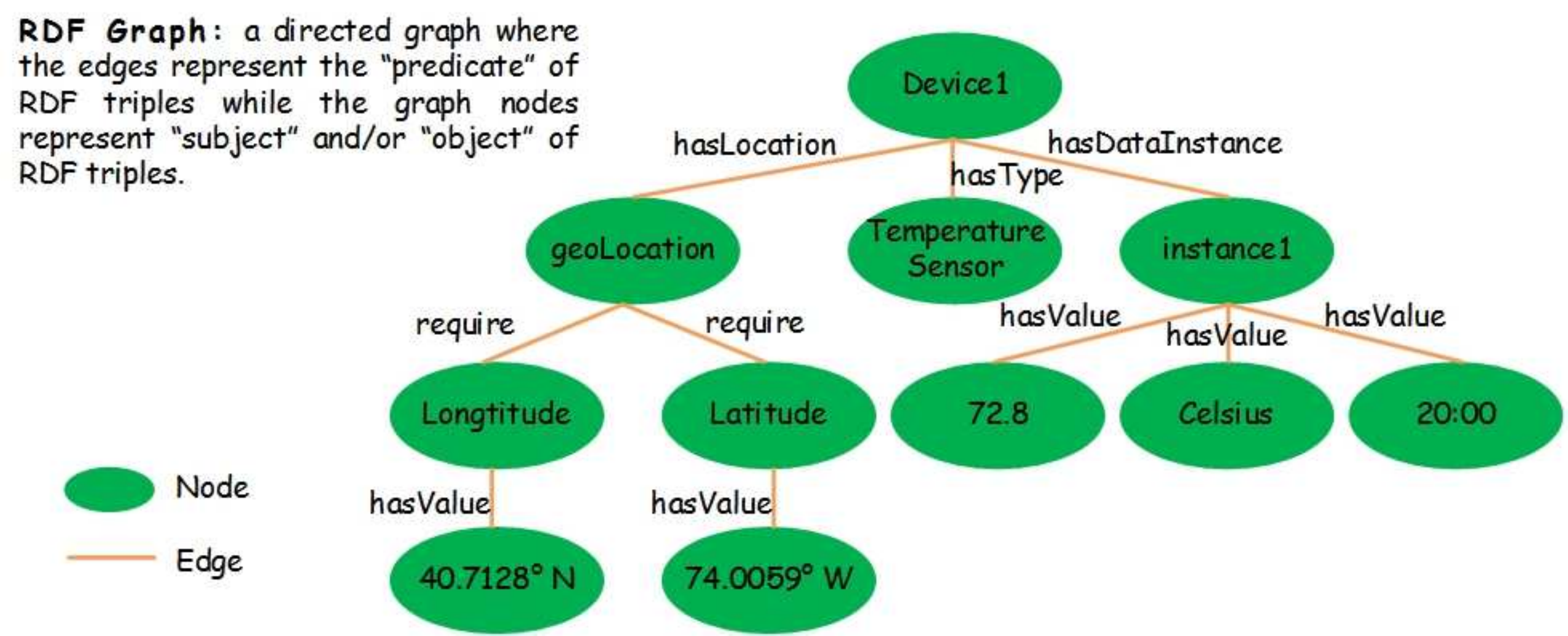}
	\caption{One example of semantic data represented by RDF graph.}	
	\label{fig:RDF_graph}	
\end{subfigure}
\caption{Semantic data.}
\label{fig:semantics}
\end{figure*}

Semantics is a method to provide a common format for annotating data such that the new semantic data can be easily found, shared, reused, and combined by machines \cite{semantic_web}. Introducing semantics into the IoT system can improve the interoperability among different IoT devices. In this section, we will illustrate how to leverage semantics in the MEIoT architecture.

All the vocabularies applied in the semantic IoT are defined in different ontologies. An ontology can be considered as a dictionary, which defines the concepts of all the vocabularies used in a specific domain as well as the relationships among different concepts. The ontologies are normally predefined and available for inference and reference. Based on these ontologies, IoT data can be represented as the machine-readable metadata format, such as RDF (Resource Description Framework) triples \cite{RDF}. An RDF triple contains three components: subject, predicate, and object. Fig. \ref{fig:semantic_data} shows one example of semantic data (corresponding to the structured data shown in Fig. \ref{fig:structured_data}) represented by RDF triples. The first two lines define two ontologies (whose URIs are "http://example.org" and "http://rdf.org", respectively) that have been applied by the following semantic data and Line$\#$3--Line$\#$11 are the semantic data represented by RDF triples. For instance, the RDF triple in Line$\#$3 depicts that the device type is a "TemperatureSensor". Note that "TemperatureSensor" is defined in the ontology, whose URI is "http://example.org", and thus the concept of "TemperatureSensor" can be retrieved from the URI of "http://example.org/TemperatureSensor". Similarly, the concept of "Type" in Line$\#$3 can be obtained from the URI of "http://rdf.org/Type". Therefore, if a device cannot understand the vocabularies applied in the semantic data, it can retrieve the corresponding concepts from related ontologies. 

%\begin{figure}[!htb]
%	\centering	
%	\includegraphics[width=1.0\columnwidth]{semantic_data.eps}
%	\caption{On example of semantic data represented by RDF triples.}	
%	\label{fig:semantic_data}	
%\end{figure}

%\begin{figure}[!htb]
%	\centering	
%	\includegraphics[width=1.0\columnwidth]{RDF_graph.eps}
%	\caption{One example of semantic data represented by RDF graph.}	
%	\label{fig:RDF_graph}	
%\end{figure}

\begin{figure*}[!htb]
\begin{minipage}[t]{1\linewidth}
	\centering	
	\includegraphics[width=1.0\columnwidth]{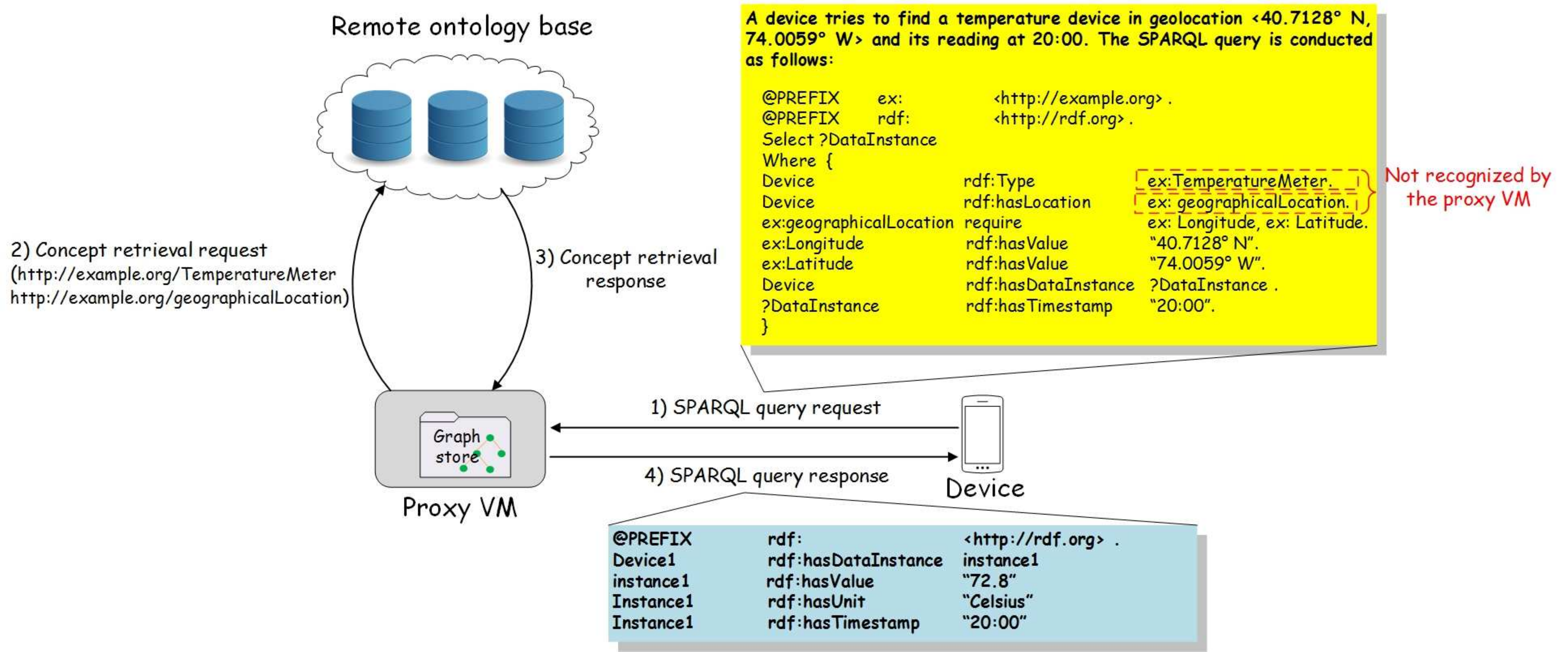}
	\caption{One example to illustrate how a proxy VM conducts SPARQL query.}	
	\label{fig:semantic_query}
\end{minipage}	
\end{figure*}

It is worth to note that RDF triples are normally stored as RDF graphs, where the edges represent the ``predicate'' of RDF triples and the nodes represent ``subject'' and/or ``object'' of RDF triples. Fig. \ref{fig:RDF_graph} shows a RDF graph related to the semantic data in Fig. \ref{fig:semantic_data}. The reason for applying RDF graphs to store semantic data is to speed up the search over a large volume of semantic data. SPARQL Protocol and RDF Query Language (SPARQL) is a query language and protocol for RDF graphs. The syntax of SPARQL is detailed in \cite{SQARQL}. Fig. \ref{fig:semantic_query} shows an example that a device (which could be a proxy VM, an application VM, or a smart user equipment with semantic capability) sends a query to the proxy VM (which contains the RDF graph shown in Fig. \ref{fig:RDF_graph}) in finding the reading of a temperature device in geolocation <40.7128$^o$ N, 74.0059$^o$ W> at 20:00\footnote{Note that a device can broadcast the query to a set of proxy VMs (e.g., all the proxy VMs within a cloudlet) to search for the requested content. Alternatively, before sending the query, a device can first discover a specific proxy VM (that may contain the requested content) by sending a proxy VM discovery request to a Resource Directory (RD) \cite{CoAP:RD}, which acts like a DNS server containing the IDs (e.g., URIs or IP addresses) and the context information of proxy VMs. The RD would respond to the device with the ID(s) of the qualified proxy VM(s) accordingly.}. The procedure is specified as follows: 1) The device would send a SPARQL query request to the proxy VM. Note that the device may not apply the same vocabularies used by the proxy VM, and thus the proxy VM may not understand the SPARQL query from the device. For instance, as shown in Fig. \ref{fig:semantic_query}, ``TemperatureMeter'' and ``geographicalLocation'' are not applied by the proxy VM, which uses ``TemperatureSensor'' and ''geoLocation'' to annotate the corresponding context of IoT data. 2) After receiving the SPARQL query, the proxy VM would send a concept retrieval request to the remote ontology base\footnote{The ontology base is considered as a repository to store ontologies and provides APIs for different devices to access these ontologies.} in order to obtain the concepts of those vocabularies. In this example, the proxy VM would sent the concept retrieval request to the URIs of ``http://example.org/TemperatureMeter'' and ``http://example.org/geographicalLocation''. 3) The remote ontology base would respond to the proxy VM with the concepts of the vocabularies. In this example, the response message could be ``$TemperatureMeter$ isSameAs $TemperatureSensor$'' and ``$geographicalLocation$ isSameAs $geoLocation$''. 4) After receiving the concept retrieval response, the proxy VM would conduct the search over its local RDF graph store and return back the results in response to the SPARQL query from the device. 

\subsubsection{Hierarchical ontology base in MEIoT}
Normally, the ontology base contains all the common vocabularies that are applied in the IoT system and is placed in the remote data center for global access. Proxy VMs need to frequently interact with the ontology base via the mobile core network in order to convert the raw IoT data into semantic data and respond to SPARQL queries. This may significantly increase the traffic load of the mobile core network and the response time in conducting SPARQL queries. 

To reduce the traffic load of the mobile core network, we propose to construct a hierarchical ontology base structure in MEIoT by placing redundant ontologies at the mobile network edge. Specifically, there are three levels of ontology bases: proxy VM ontology base, cloudlet ontology base, and global ontology base. A proxy VM ontology base is placed in each proxy VM. It stores the concepts of all the vocabularies that are used by the proxy VM's registered IoT devices. For example, if a temperature sensor is registered to a proxy VM, then the proxy VM should download the concepts of all the vocabularies related to the temperature sensor from the global ontology base and store them in its proxy VM ontology base. A cloudlet ontology base is deployed in each cloudlet. It stores the concepts of vocabularies related to some applications, which are mainly determined by the location of the cloudlet. For instance, if a cloudlet is deployed in the residential area, its ontology base should store the concepts of the vocabularies related to the smart home application. A global ontology base contains all the common vocabularies and is normally located in a remote data center.

\subsection{Challenge 2: Inefficient access control in IoT}

IoT data streams normally contain users' personal information (such as users' location traces, health status, etc.), and thus it is important to provide an efficient privacy-preserving solution for the system to identify a device's permission to access the corresponding data. One of the most common Access Control (AC) models that has been applied in the IoT system is the AC list \cite{Mahalle:2013:IAC,Zhang:2016:ACI}, where access rights/policies are listed in IoT devices. For example, if a mobile phone tries to access the data provided by a temperature sensor node, the mobile phone should send a data access request containing its identification and/or contextual information (e.g., IP address and/or geolocation of the mobile phone) to the temperature sensor, which checks its AC policies to see if the requested device has the privileges to manipulate (i.e., retrieve, update, delete, observe, etc.) the corresponding data. 

The AC list method is, however, not efficient and scalable. 1) Different devices may use different identification strategies; for instance, a smart TV may use its manufacturer model number, a laptop may apply a product key of its operating system, and a temperature sensor may use its IP/MAC address to identify themselves. It is difficult to build and maintain a complete AC list to enable/disable authorized/unauthorized devices (which may adopt different identification strategies) in accessing the corresponding data. 2) An AC list is not automatically generated, i.e., an IoT device owner may need to manually setup/update an AC list. 3) IoT devices are mostly resource constrained, and thus cannot store and maintain a huge volume of AC lists. Hence, it is necessary to design an efficient and scalable AC mechanism to automatically generate AC policies for each device with the consideration of the resource constrained features of IoT devices. 

\subsection{Solution 2: Efficient access control mechanism based on semantic social IoT}

\begin{figure*}[!htb]
\begin{minipage}[t]{1\linewidth}
	\centering	
	\includegraphics[width=0.9\columnwidth]{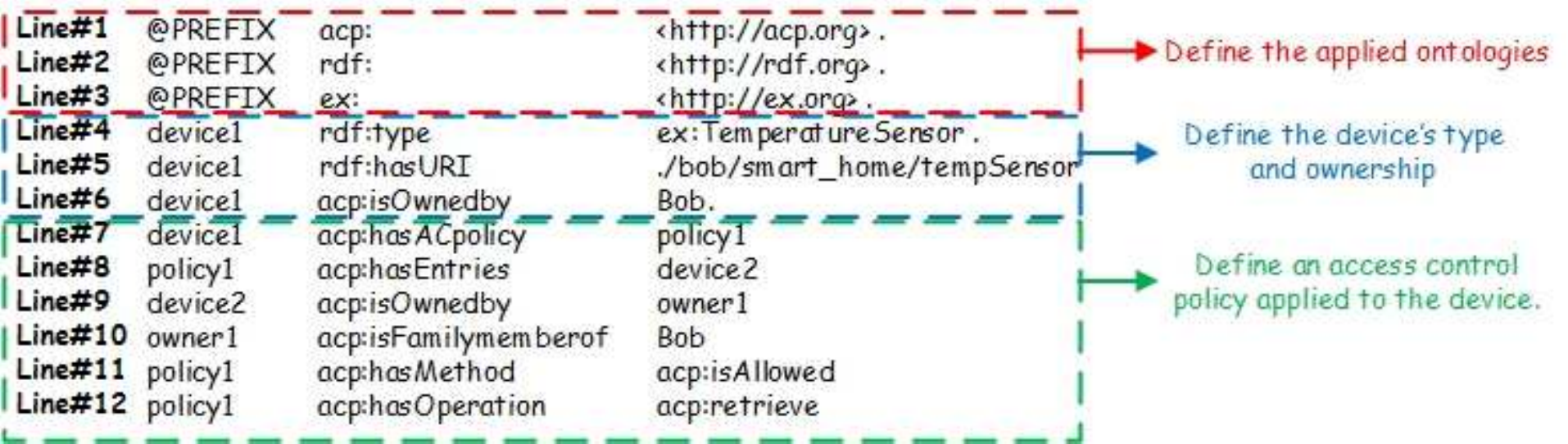}
	\caption{One example to illustrate an access control policy in social IoT.}	
	\label{fig:access_control_policy}
\end{minipage}	
\end{figure*}

Social IoT (SIoT) \cite{SIoT,Atzori:2014:FSO} is to combine the IoT system with the concept of ``human social network'' such that IoT devices are capable of building social relationships with other IoT devices without human involvement. It is worth to note that the relationships among IoT devices do not just only rely on the human relationships, but also depend on the contextual information of those IoT devices. Farris \emph{et al.} \cite{Farris:2015:SVO} summarized the relationships among IoT devices as follows: 1) \emph{parental relationship}: devices produced by the same manufacturer; 2) \emph{co-location relationship}: devices are located in the same places; 3) \emph{co-work relationship}: devices collaborate together to implement the same IoT service; 4) \emph{ownership relationship}: devices belong to the same user; 5) \emph{social relationship}: devices communicate with each other because of the relationships among their owners. %In addition, we add one more relationship, i.e., 6) \emph{server-client relationship}: when an application is installed in a proxy VM, the installed application has the right to access the corresponding data in the proxy VM. Thus, the installed application and the proxy VM follows a server-client relationship. For instance, if a terrorist detection application is installed in an proxy VM, the application can access the data related to the photos/videos in the proxy VM, generate the metadata (i.e., the contextual information of the matched photos/videos), and send them to the application VM. Thus, terrorist detection application and the proxy VM follows a server-client relationship, where the terrorist detection application is considered as a client (which always retrieves photos/videos from the server) and the proxy VM is considered as a server (which always provides related data to the client).

It is thus efficient to set up the AC policies based on the social relationships among different IoT devices. For instance, if the temperature sensor in Bob's smart home can be accessed by the devices, which are owned by Bob's family members, then we can build an AC policy based on the ownership relationship. The corresponding AC policy can be specified as shown in Fig. \ref{fig:access_control_policy}, where the AC policies are represented by RDF triples. Here, Line$\#$1--Line$\#$3 define the applied ontologies; Line$\#$4--Line$\#$6 imply the type, URI, and ownership of the temperature sensor; Line$\#$7--Line$\#$12 depict an access control policy, i.e., policy1, associated to the device. An access control policy should define two components. 1) What operations can/cannot be performed on the device. In this example, Line$\#$11--Line$\#$12 describe the retrieve operation that can be conducted. 2) Who can/cannot perform these operations. In this example, Line$\#$7--Line$\#$10 define the family members of Bob (the owner of the device) that can retrieve the temperature value of the device. Note that applying semantics to annotate access control policies is to improve the interoperability among IoT devices and social networks. 

Since most of IoT devices are resource constrained, it is not efficient to enable IoT devices to conduct access control. Instead, the access control functionalities are outsourced to proxy VMs, i.e., proxy VMs would maintain their registered devices' AC policies and handle data access requests from devices to determine whether they have the privileges to access (i.e., retrieve, update, delete, observe, etc.) the corresponding data. To better illustrate the procedure of a proxy VM in conducting access control, we provide a simple scenario in which Alice's mobile phone tries to retrieve the temperature value sensed by the temperature sensor (whose AC policy is specified in Fig. \ref{fig:access_control_policy}) located in Bob's smart home.  The whole procedure comprises the following five steps.
\begin{figure*}[!htb]
\begin{minipage}[t]{1\linewidth}
	\centering	
	\includegraphics[width=1.0\columnwidth]{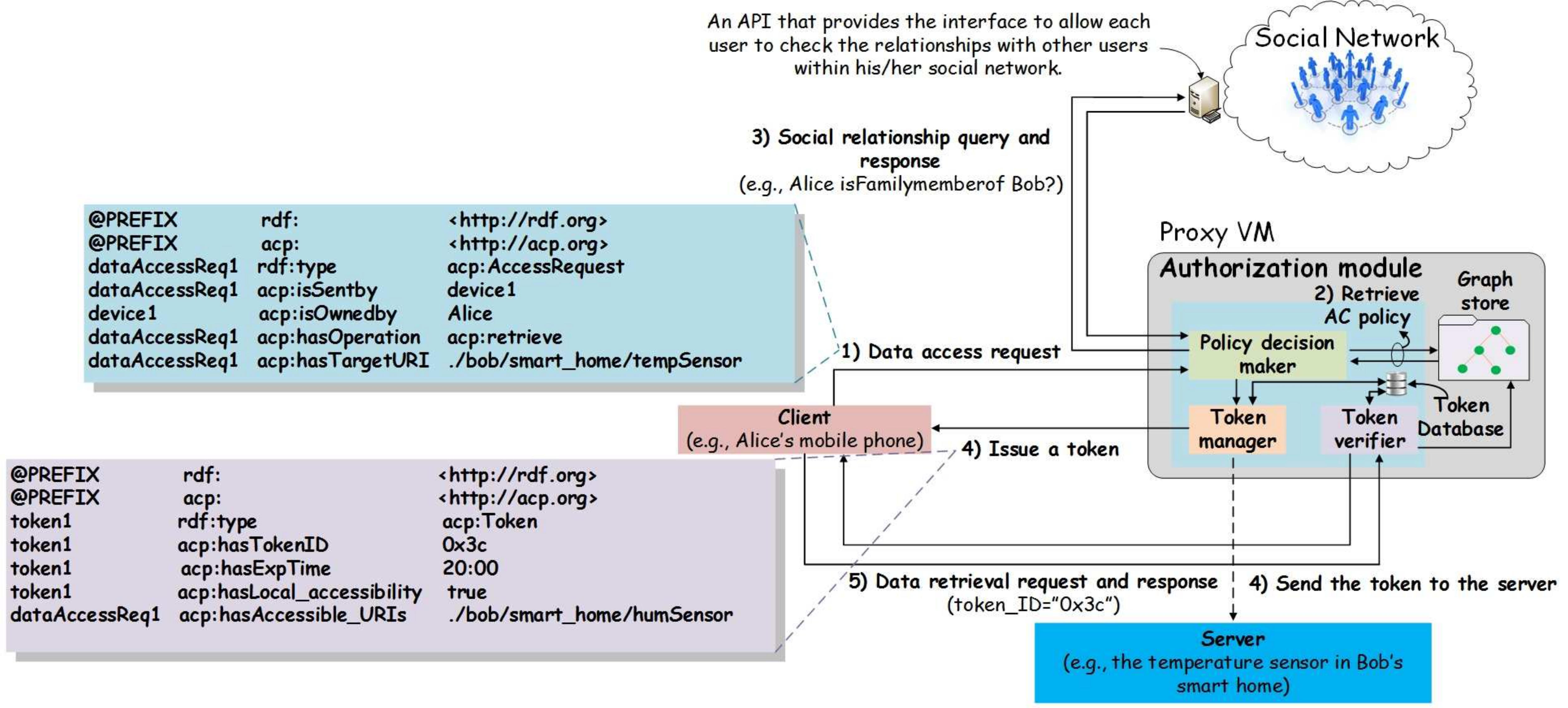}
	\caption{The procedure of a proxy VM in conducting access control.}	
	\label{fig:access_on_proxyVM}
\end{minipage}	
\end{figure*}
\begin{enumerate}
\item{A client (i.e., Alice's mobile phone) sends a data access request to the proxy VM (which is associated to the temperature sensor). The request should contain the information on which operation is requested to be performed on which device as well as the contextual information of the client (e.g., the ownership and location of the client). For example, as shown in Fig. \ref{fig:access_on_proxyVM}, the client sends a data access request to the proxy VM in retrieving the data provided by the temperature sensor (which is identified by the URI of ``./bob/smart$\_$home/tempSensor''). The request contains the contextual information of the client (e.g., the ownership) to identify itself. Note that the client could be any device (e.g., a mobile phone, a proxy VM, an application VM, etc.) and the data access request could be sent directly to the temperature sensor (rather than the proxy VM), which then relays this request to its proxy VM for conducting access control.}
\item{After the proxy VM receives the request, the Policy Decision Maker (PDM) in the proxy VM's authorization module would try to obtain the corresponding AC policies related to the temperature sensor by performing a semantic query over its local graph store. After obtaining the AC policies, PDM would check if the client has the privilege to access the data provided by the temperature sensor. PDM could immediately allow/deny the request if the contextual information of the client satisfies/does not satisfy the AC policy. Also, PDM could ask the client to provide more contextual information to identify itself. In this example, the AC policy (as shown in Fig. \ref{fig:access_on_proxyVM}) specifies that the data can be retrieved by the devices owned by Bob's family members. Thus, the proxy VM needs to check the relationship between Alice and Bob.}
\item{PDM sends a social relationship query to the social network to identify the social relationship between the owner of the temperature sensor and the owner of the client. In this example, the family membership between Alice and Bob\footnote{Note that the identities of owners, i.e., ``Alice'' and ``Bob'', are encrypted at the device side and will be decrypted and identified at the social network side} should be confirmed in the Bob's social network. As a response, the social network would return a positive/negative message to confirm/deny the relationship.}
\item{If the social relationship is confirmed, PDM would ask the token manager in the proxy VM's authorization module to issue a token to the client. A token is considered as a permission that allows its holder to access the data within a valid time interval \cite{Bodriagov:2011:EPP}. A token should contain the $hasTokenID$, $hasExpTime$, $hasLocal\_accessibility$, and $hasaccessible\_URI$ attributes, where $hasTokenID$ implies the ID of the issued token; $hasExpTime$ indicates the expiration time of this token. After the expiration, the client cannot use this token to access the corresponding data and the identity of the client should be reevaluated by PDM. $hasLocal\_accessibility$ indicates whether the client can access the data in the temperature sensor (rather than the data stored in the temperature sensor's proxy VM). If $hasLocal\_accessibility="true"$, the proxy VM should send the token ID to the temperature sensor such that the client can access the data in the temperature sensor by applying the same token ID. $hasaccessible\_URI$ provides a list of URIs of other devices, which share the same access control policy, that could be accessed by the client. For example, if the client can retrieve the data of the temperature sensor, it can also retrieve the data of the humidity sensor in the smart home.}
\item{After receiving the token, the client can retrieve the data by sending a data retrieval request containing the corresponding $token\_ID$ to the proxy VM. The token verifier in the proxy VM's authorization module would verify the $token\_ID$ and respond to the client with the temperature value (which is retrieved from the local graph store) if the $token\_ID$ has the privilege to retrieve the data from the server.}
\end{enumerate}  

The proposed Semantic SIoT provides a flexible and efficient access control mechanism, which is to enforce access control policies based on social relationships among different devices. Also, devices, which outsource their access control functionalities to their proxy VMs, may reduce their energy consumption and accelerate the access control precess.

\subsection{Challenge 3: Mobility problem in MEIoT}
In the MEIoT architecture, each IoT device is associated to a specific proxy VM located in the nearby cloudlet. The proxy VM helps its registered IoT devices to share, store, and process their generated data with low End-to-End (E2E) delay. This can substantially reduce the energy consumption of IoT devices and accelerate the process of IoT data sharing and analysis. However, some of the IoT devices (e.g., smart phones) are mobile and statically placing their proxy VMs in the original cloudlets may not benefit the process of IoT data sharing and analytics, i.e., mobile devices need to upload their generated data streams to their proxy VMs via the mobile core network; this may increase the traffic load of the mobile core network as well as the E2E delay between mobile devices and their proxy VMs. Note that, as mentioned in Section \ref{sec:Introduction}, delay is a critical factor that affects the QoS of many IoT applications. Thus, it is important to keep the E2E delays between mobile devices and their proxy VMs low in order to satisfy the stringent delay requirement of these IoT applications.

\subsection{Solution 3: Latency aware proxy VM migration among cloudlets in MEIoT}
\begin{figure}[t]
	\centering	
	\includegraphics[width=1.0\columnwidth]{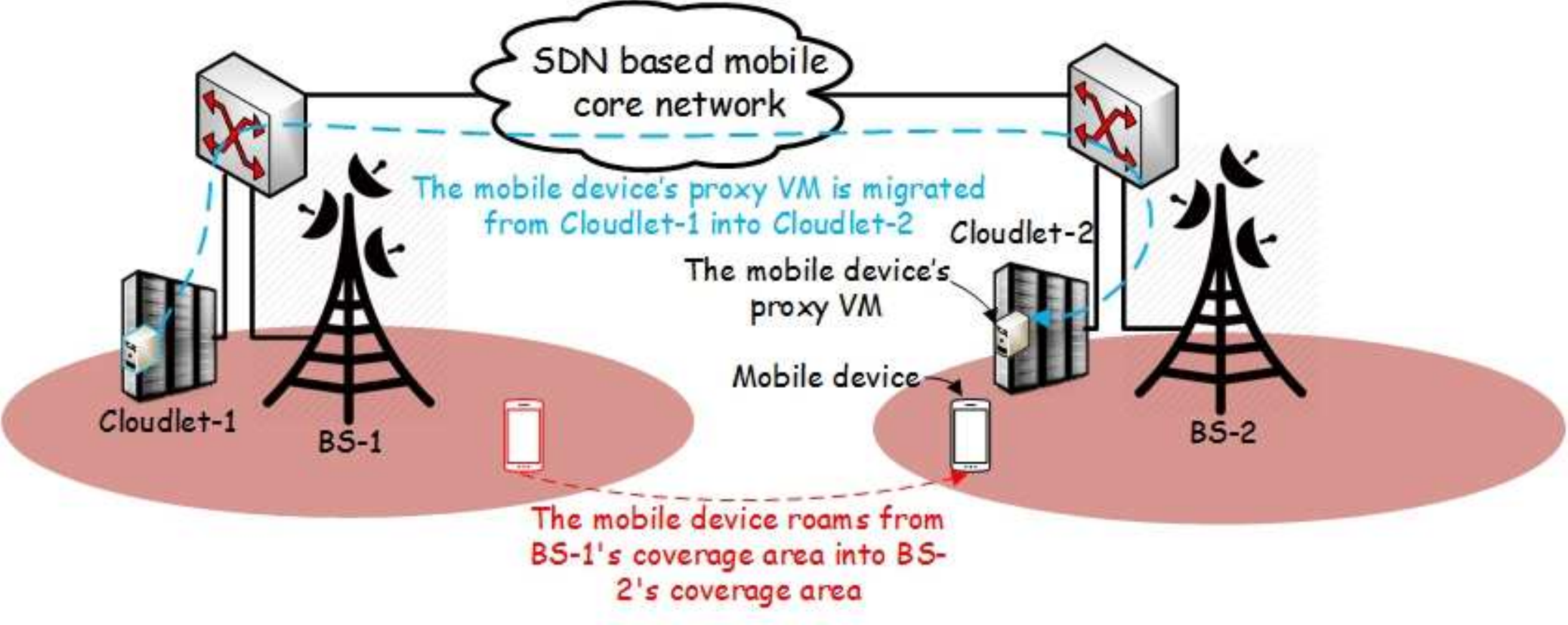}
	\caption{Dynamic proxy VM migration among different cloudlets.}	
	\label{fig:dyamnic_migration}	
\end{figure}
In order to keep the E2E delays between mobile devices and their proxy VMs low, the location proxy VMs can be dynamically changed based on the locations of the mobile devices \cite{Sun:2016:PRIMAL}. As shown in Fig. \ref{fig:dyamnic_migration}, if a mobile device roams from BS-1's coverage area into BS-2's coverage area, its proxy VM can be migrated to cloudlet-2. Definitely, the E2E delay between the mobile device and its proxy VM in cloudlet-2 is lower than that between the mobile device and its proxy VM in cloudlet-1. However, each cloudlet has the limitation to accommodate the number of proxy VMs, i.e., a cloudlet may not have enough space to host all the proxy VMs of the local mobile devices of the cloudlet. Thus, it is nontrivial to determine the locations of proxy VMs to minimize the total average E2E delay between mobile devices and their proxy VMs while jointly considering the capacity limitation of each cloudlet.

Denote $\bm{\mathcal{I}}$, $\bm{\mathcal{J}}$ and $\bm{\mathcal{K}}$ as the set of mobile devices, BSs, and cloudlets, respectively. Denote $x_{ik}$ as the binary variable to indicate whether mobile device $i$'s proxy VM is in cloudlet $k$ (i.e., $x_{ik}=1$) or not (i.e., $x_{ik}=0$), where $i \in \bm{\mathcal{I}}$ and $k \in \bm{\mathcal{K}}$. Meanwhile, let $\tau_{jk}$ be the average E2E delay between BS $j$ and cloudlet $k$, where $j \in \bm{\mathcal{J}}$. Note that the value of $\tau_{jk}$ can be measured by the SDN controller periodically \cite{Nadembega:2015:DMP,Adrichem:2014:ONM}. We consider the E2E delay between a mobile device and its proxy VM as the E2E delay between the mobile device's associated BS and the cloudlet (which hosts the mobile device's proxy VM)\footnote{The E2E delay between a mobile device and its proxy VM comprises the E2E delay between the mobile device and its associated BS, the E2E delay between the associated BS and the cloudlet (which hosts the mobile device's proxy VM), and the E2E delay within the cloudlet. Yet, optimizing the location of the proxy VM cannot change the E2E delay between the mobile device and the BS; meanwhile, the E2E delay within the cloudlet is negligible. Hence, we consider the E2E delay between the mobile device and its proxy VM as the E2E delay between the mobile device's associated BS and the cloudlet for the rest of the paper.}.  Moreover, denote $\psi_{ij}$ as the location indicator to imply whether mobile device $i$ is in BS $j$'s coverage area (i.e., $\psi_{ij}=1$) or not (i.e., $\psi_{ij}=0$). Suppose each proxy VM has the same configuration (i.e., the same amount of CPU, memory, and hard disk resource reservations) and denote $\phi_k$ as the capacity of cloudlet $k$, i.e., the maximum number of proxy VMs that cloudlet $k$ can accommodate. Then, we formulate the latency aware proxy VM migration problem as follows: 

\begin{align}
&\mathop {\arg \min }\limits_{\bm{\mathcal{X}}} \sum\limits_{i \in \bm{\mathcal{I}}} {\sum\limits_{j \in \bm{\mathcal{J}}} {\sum\limits_{k \in \bm{\mathcal{K}}} {{\psi _{ij}}{\tau _{jk}}{x_{ik}}} } }, \\
s.t.\ \ \ &\sum\limits_{k \in \bm{\mathcal{K}}} {{x_{ik}}}  = 1,\ \forall i \in \bm{\mathcal{I}},\label{ct_1}\\
&\sum\limits_{i \in \bm{\mathcal{I}}} {{x_{ik}}}  \le {\phi _k},\ \forall k \in \bm{\mathcal{K}},\label{ct_2}\\
&\forall k \in \bm{\mathcal{K}},{x_{ik}} \in \left\{ {0,1} \right\},\ \forall i \in \bm{\mathcal{I}}. \label{ct_3}
\end{align}
The objective is to minimize the total E2E delay between all the mobile devices and their proxy VMs, where $\bm{\mathcal{X}}=\left\{ {{x_{ik}}\left| {i \in \bm{\mathcal{I}},k \in \bm{\mathcal{K}}} \right.} \right\}$. Constraint \eqref{ct_1} imposes each mobile device's proxy VM to be placed in one cloudlet. Constraint \eqref{ct_2} imposes the number of proxy VMs hosted by a cloudlet not to exceed the capacity of the cloudlet. Constraint \eqref{ct_3} means $x_{ik}$ is a binary variable. Note that the latency aware proxy VM migration problem is an integer binary programming problem, and thus we can apply the commercial solver, i.e., CLPEX, to solve the problem.

\subsection{Challenge 4: Energy inefficiency in MEIoT}

Maintaining a large number of cloudlets incurs a huge operational expenditure to the cloudlet provider by paying an expensive energy bill to the on-grid energy suppliers. Green energy can be leveraged to reduce the operational expenditure. Specifically, each cloudlet is powered by both green energy and on-grid energy. Green energy is generated from renewal resources (e.q., solar, wind, geothermal, etc.) and is considered as a ``free'' energy supply for the cloudlet provider; on-grid energy is pulled from the smart grid and is considered as a backup energy supply for each cloudlet (i.e., a cloudlet would consume on-grid energy only if there is no residual green energy to power the cloudlet). Detailed descriptions of the green cloudlet system can be found in \cite{Sun:2017:GCN}. It is worth to note that green energy is discouraged to be ``banked'' since many disadvantages have been proved in storing the superfluous green energy in batteries \cite{Goiri:2015:SVO}. Therefore, if green energy is not fully utilized by cloudlets in the current time slot, it will be wasted.

Green energy generation exhibits spatial dynamics, i.e., green energy generated in different cloudlets may vary \cite{Ansari:2017:GMN}. Also, different cloudlets may host different number of proxy VMs, and thus the energy demand exhibits spatial dynamics as well, i.e., different cloudlets may have different energy demands. The spatial dynamics of green energy generation and energy demands invoke the problem of unbalanced energy gap\footnote{Energy gap of a cloudlet is defined as the difference between the energy demand and the green energy generation of the cloudlet. Positive energy gap means the generated green energy cannot satisfy the energy demand of the cloudlet and negative energy gap implies the cloudlet has superfluous green energy to meet its energy demand.} among cloudlets, thus resulting in inefficient energy utilization and increasing the operational costs for the cloudlet provider.

\subsection{Solution 4: Energy driven proxy VM migration among cloudlets in MEIoT}

In order to balance the energy gap among cloudlets, we propose to migrate proxy VMs from the cloudlets with negative energy gap into the cloudlets with positive energy gap \cite{Sun:2015:GEA}. Specifically, the energy demand of cloudlet $k$ equals to the sum of the energy consumption of working PMs\footnote{A PM is said to be a working PM if it hosts at least one proxy VM.}. The energy consumption of each working PM is modeled as 
\begin{equation}
{p^{PM}} = \Delta T\left( {{\rho ^s} + \alpha \mu } \right),
\end{equation} 
where $\Delta T$ is the duration of one time slot, $\rho^s$ is the static power consumption of a PM (the power consumption of a PM when it is in the standby mode), and $\alpha \mu$ is the dynamic power consumption of a PM. Here, $\mu$ is the CPU utilization of a PM, which equals to the sum of the CPU utilization of its hosted proxy VMs and $\alpha$ is the power coefficient that maps the CPU utilization into power consumption. If there are $N_k$ working PMs in cloudlet $k$, we can derive the energy demand of cloudlet $k$ as
\begin{equation}
{p_k} = \Delta T\left( {{N_k}{\rho ^s} + \alpha \sum\limits_{i \in \bm{\mathcal{I}}} {{\mu _i}{x_{ik}}} } \right),
\end{equation} 
where ${N_k}{\rho ^s}$ and $\alpha \sum\limits_{i \in \bm{\mathcal{I}}} {{\mu _i}{x_{ik}}}$ are total static and dynamic power consumption of all the working PMs in cloudlet $k$, respectively. Here, $\mu _i$ indicates the CPU utilization of mobile device $i$'s proxy VM. Note that the value of $N_k$ is determined by the number of proxy VMs in cloudlet $k$. If each PM in a cloudlet can host $\epsilon$ number of proxy VMs, then ${N_k} = \left\lceil {\frac{{\sum\limits_{i \in \bm{\mathcal{I}}} {{x_{ik}}} }}{\epsilon}} \right\rceil$, where $\left\lceil  \bullet  \right\rceil$ is the ceiling function. By approximating ${N_k} \approx \frac{{\sum\limits_{i \in \bm{\mathcal{I}}} {{x_{ik}}} }}{\epsilon}$, we have
\begin{equation}
{p_k} = \Delta T\sum\limits_{i \in \bm{\mathcal{I}}} {\left( {\frac{{{\rho ^s}}}{\varepsilon } + \alpha {\mu _i}} \right){x_{ik}}}. 
\end{equation}
Thus, the energy driven proxy VM migration problem can be formulated as follows:
\begin{align}
&\mathop {\arg \min }\limits_{\bm{\mathcal{X}}} \sum\limits_{k \in \bm{\mathcal{K}}} {\max \left\{ {\Delta T\sum\limits_{i \in \bm{\mathcal{I}}} {\left( {\frac{{{\rho ^s}}}{\epsilon }\!+\!\alpha {\mu _i}} \right){x_{ik}}\!-\!{g_k}} ,0} \right\}}, \\
s.t.\ &\sum\limits_{j \in \bm{\mathcal{J}}} {\sum\limits_{k \in \bm{\mathcal{K}}} {{\psi _{ij}}{\tau _{jk}}{x_{ik}}} }  \le \gamma,\ \forall i \in \bm{\mathcal{I}}, \label{ct_4}\\
& Constraints\ \eqref{ct_1}, \eqref{ct_2}, and\ \eqref{ct_3},
\end{align}
where $g_k$ indicates the total amount of green energy generated by cloudlet $k$ during a time slot and $\gamma$ is an E2E delay threshold, which defines the maximum E2E delay between a mobile device and its proxy VM. The objective of the problem is to minimize the total on-grid energy consumption and Constraint \eqref{ct_4} is to guarantee the E2E delay between every mobile device and its proxy VM to be less than a pre-defined threshold $\gamma$. Note that the energy driven proxy VM migration problem is a mixed integer linear programming problem, and thus we also can use CPLEX to solve it. 

\section{Performance evaluation}
In this section, we will evaluate the performance of Latency Aware proxy VM Migration (LAM) and Energy Aware proxy VM Migration (EAM) as compared to the Static method. The Static method means proxy VMs do not change their locations after the initial deployment.

\begin{figure}[t]
	\centering	
	\includegraphics[width=0.8\columnwidth]{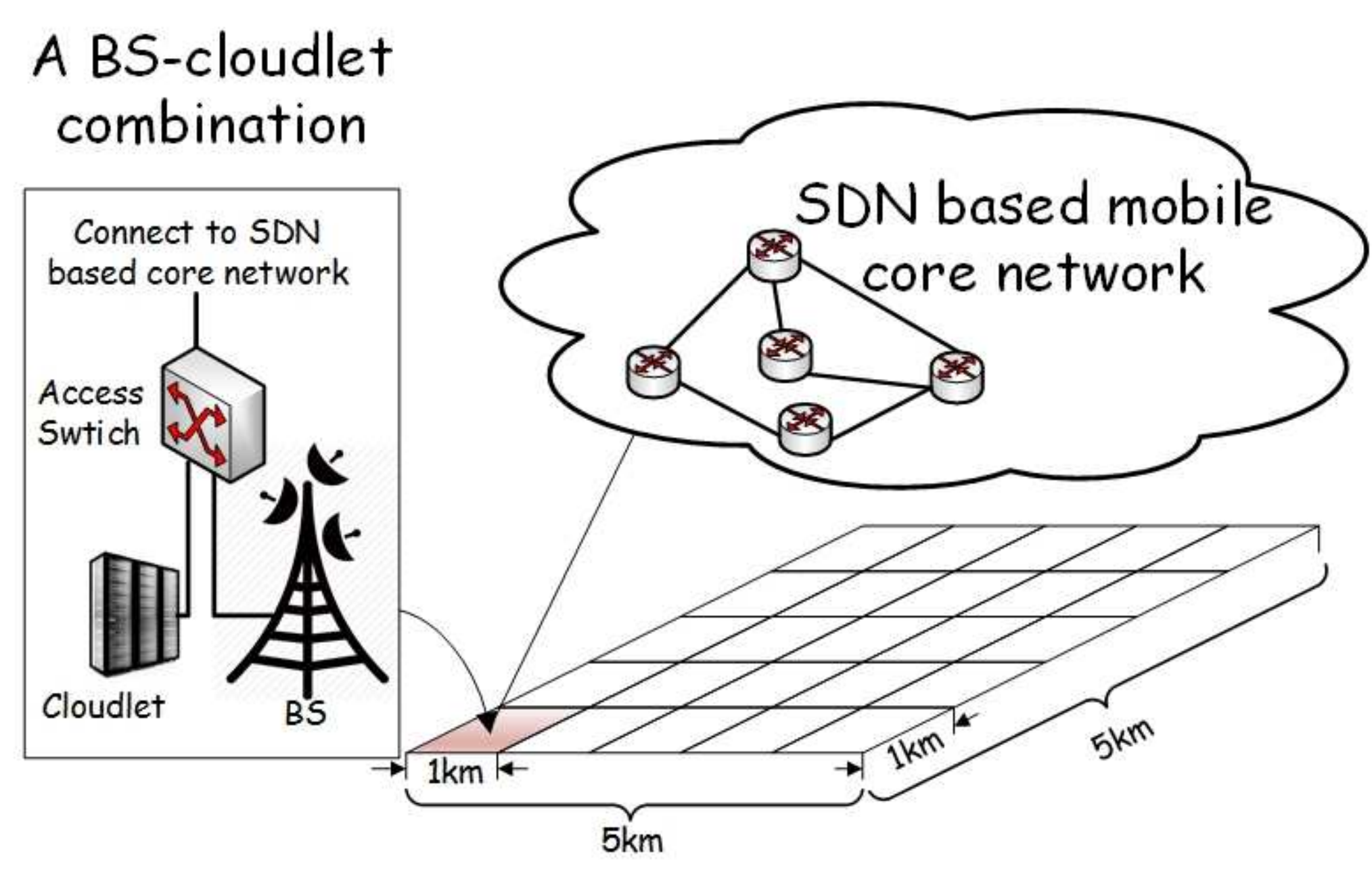}
	\caption{The network topology.}	
	\label{fig:network_topo}
\end{figure}

\begin{table}[t]
  \centering
  \caption{Values and Definitions of Parameters}
    \begin{tabular}{lll}
    \toprule
    \textbf{Parameters} & \textbf{Definition} & \textbf{Value}\\
    \midrule
    $\alpha$ & power coefficient & 0.2 $W/\%$ \\
    $\beta$ &  E2E delay offset & 10 $ms$ \\
    $\lambda$ & E2E delay coefficient & 25 $ms/km$ \\
    $\gamma$ & E2E delay threshold & 40 $ms$\\  
    $\Delta T$ & the length of a time slot & 0.5 $hour$ \\
    $\rho^s$ & static power consumption of a PM &80 $W$ \\
    $\left| \bm{\mathcal{I}} \right|$    & total number of qualified mobile devices & 632 \\
    $\left| \bm{\mathcal{J}} \right|$     & total number of BSs & 25 \\
    $\left| \bm{\mathcal{K}} \right|$     & total number of cloudlets & 25 \\
    \bottomrule
    \end{tabular}%
  \label{tab:sim_para}%
\end{table}

We set up a network with $5 \times 5$ BS-cloudlet combinations. As shown in Fig. \ref{fig:network_topo}, each BS is connected to its cloudlet via an access switch. Also, each BS/cloudlet can communicate with other BSs/cloudlets via the SDN based cellular core network and each cloudlet is powered by both green energy and on-grid energy. The amount of green energy generated in each cloudlet is considered to be the same, i.e., $\forall k \in \bm{\mathcal{K}}$, $g_k=1000W$. The radio coverage area for each BS is $1km\times1km$. In order to emulate each mobile device's movement pattern, we apply the user movement trace provided by the EveryWare Lab \cite{EveryWare_Lab}. The trace provides the users' movement in the road network of Milan in different time slots. We select a $5km \times 5km$ area of the network and monitor the movement of the qualified users\footnote{The qualified users are users who only move within the selected $5km \times 5km$ area during the monitoring period.} in different time slots during the monitoring period of six hours. We use the qualified users' movement trace to obtain the values of ${\bm{\psi }} = \left\{ {{\psi _{ij}}\left| {i \in \bm{\mathcal{I}}, j \in \bm{\mathcal{J}}} \right.} \right\}$ in different time slots. In addition, we assume that the E2E delay between a clouldet and a BS is proportional to their distance\footnote{The E2E delay between a clouldet and a BS is measured by the SDN controller in the real system.}, i.e., $\tau_{jk}=\lambda d_{jk}+\beta$, where $d_{jk}$ is the distance between BS $j$ and cloudlet $k$, and $\lambda$ and $\beta$ are the coefficient and the offset used to map distance into delay, respectively. Moreover, the CPU utilization of each mobile device's proxy VM is randomly selected between 20\% and 100\%, i.e., $\mu_i=U\left( {0.2,1} \right), \forall i \in \bm{\mathcal{I}}$. Each cloudlet contains 5 PMs and each PM can host at most 6 proxy VMs, i.e., $\epsilon=6$ and $\phi _k=30$. Other simulation parameters are listed in Table \ref{tab:sim_para}.

\begin{figure*}[!htb]
\centering
\begin{subfigure}{.4\textwidth}
  \centering
  \includegraphics[width=1.0\linewidth]{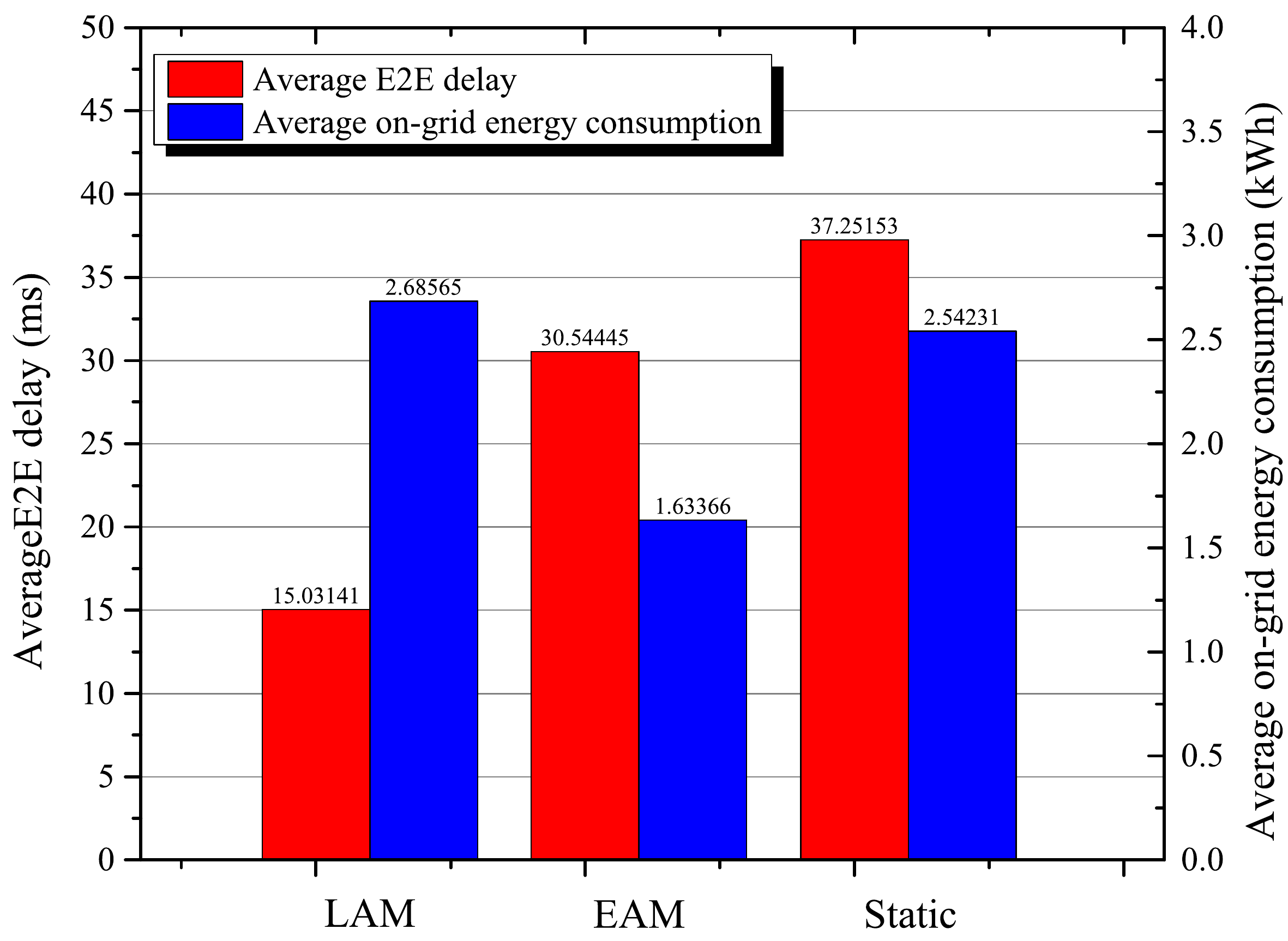}
  \caption{Average on-grid energy consumption and average E2E delay.}
  \label{fig:overall_perfm}
\end{subfigure}%
\hspace{0.2cm}
\begin{subfigure}{.4\textwidth}
  \centering
  \includegraphics[width=1.0\linewidth]{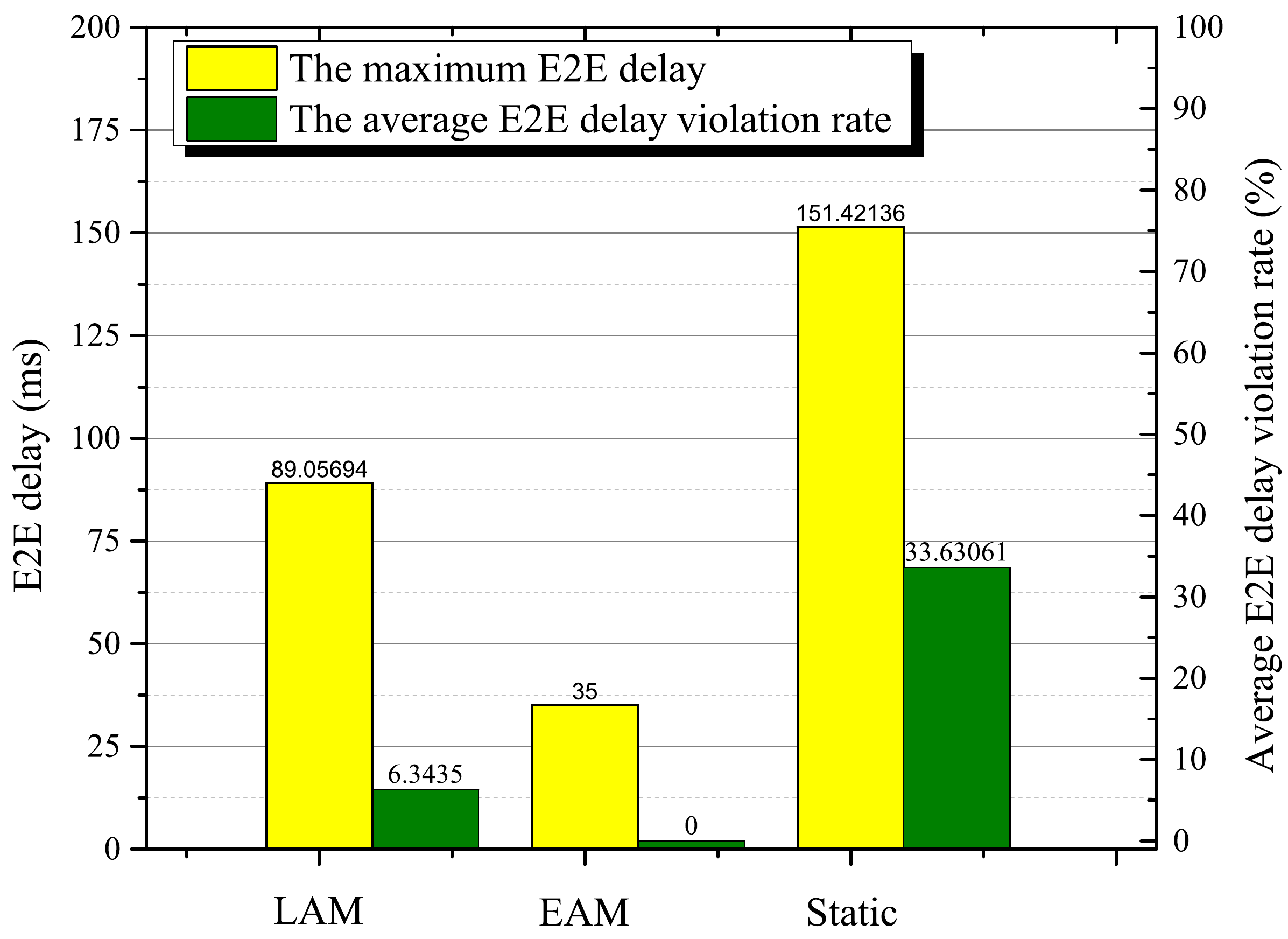}
  \caption{The maximum E2E delay and average E2E delay violation rate.}
  \label{fig:overall_perfm_SLA}
\end{subfigure}
\hspace{0.2cm}
\begin{subfigure}{.4\textwidth}
  \centering
  \includegraphics[width=1.0\linewidth]{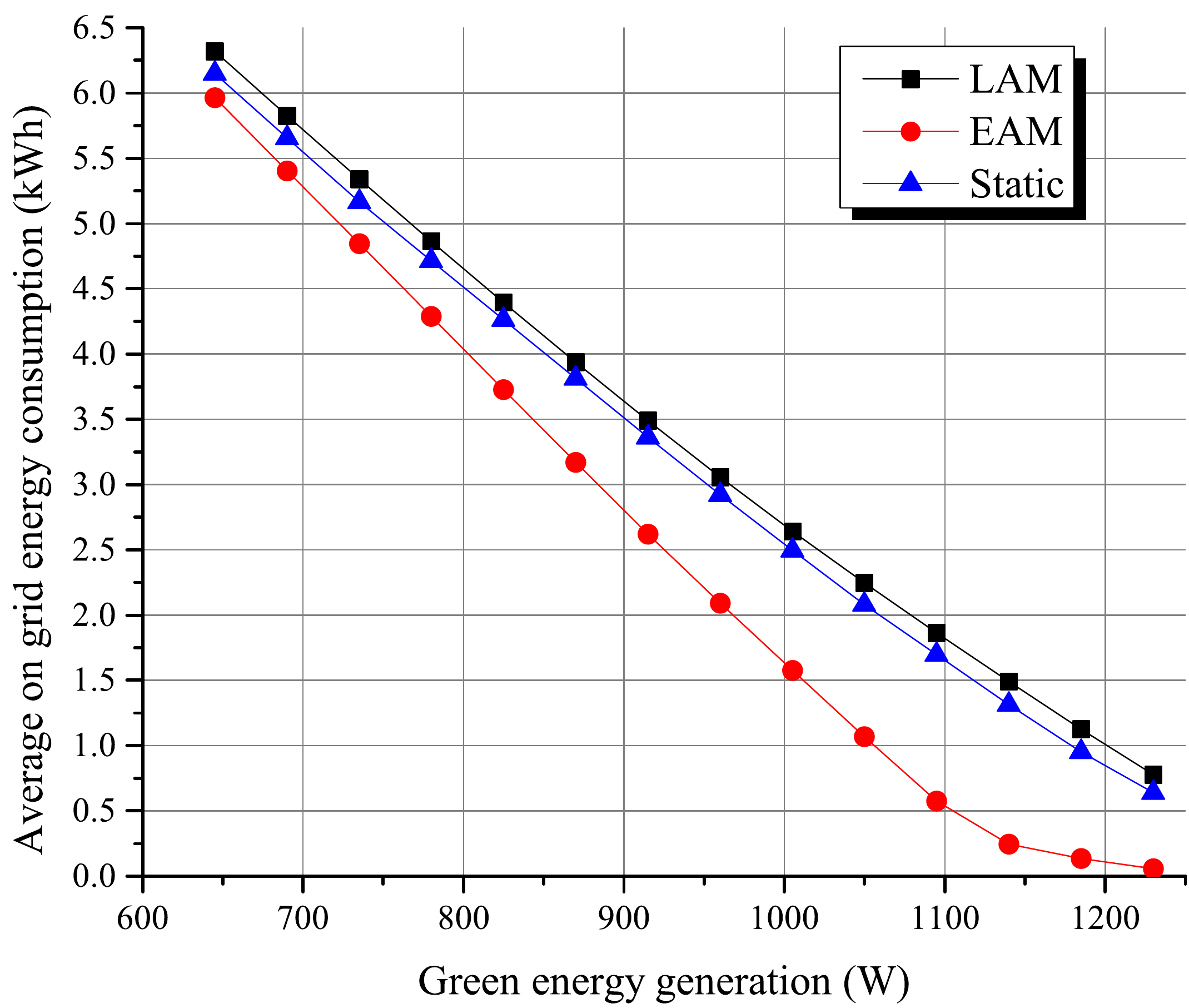}
  \caption{Average on-grid energy consumption by varying green energy generation.}
  \label{fig:perfm_green_consumption}
\end{subfigure}
\hspace{0.2cm}
\begin{subfigure}{.4\textwidth}
  \centering
  \includegraphics[width=1.0\linewidth]{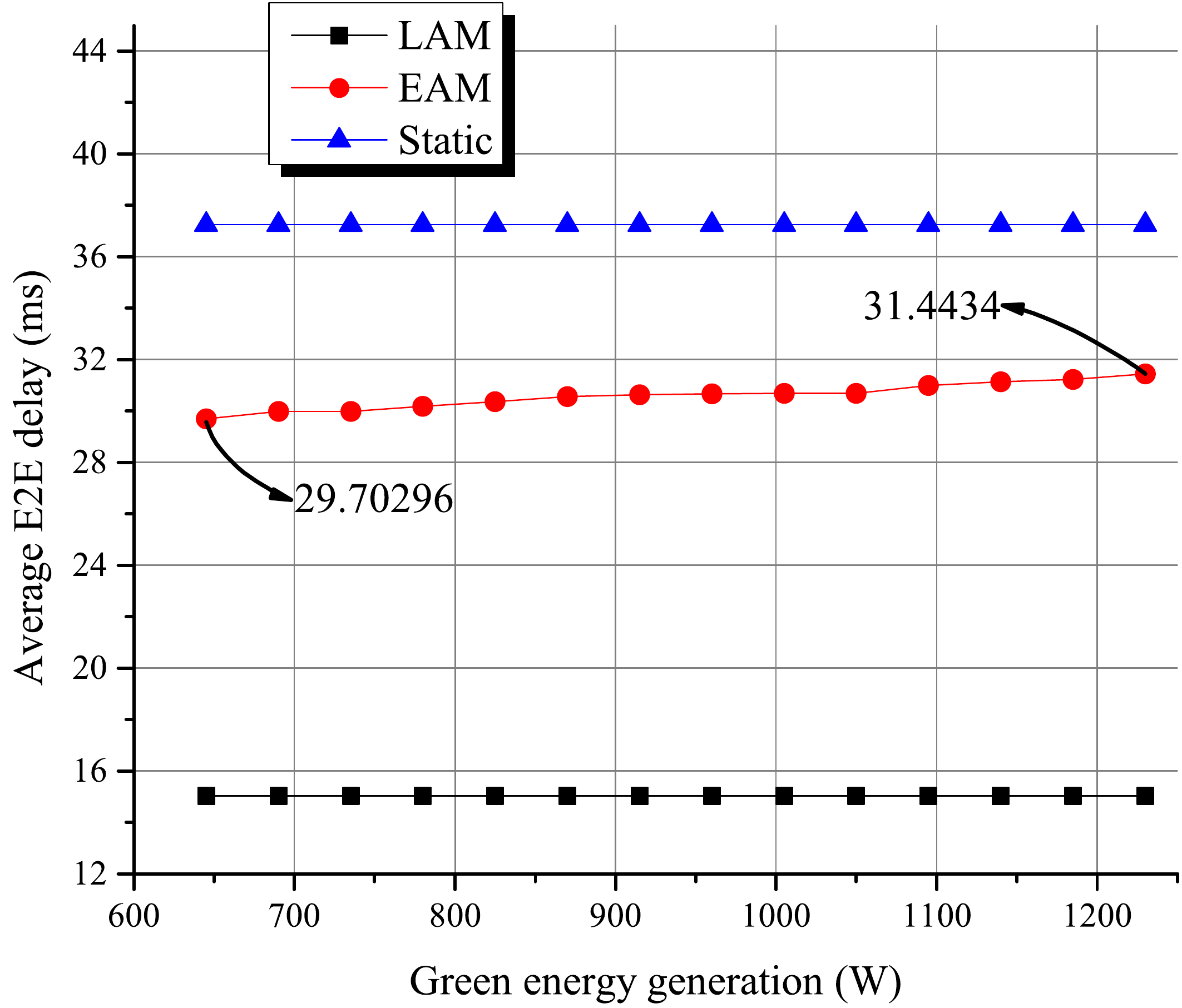}
  \caption{Average E2E delay by varying green energy generation.}
  \label{fig:perfm_green_delay}
\end{subfigure}
\hspace{0.2cm}
\begin{subfigure}{.4\textwidth}
  \centering
  \includegraphics[width=1.0\linewidth]{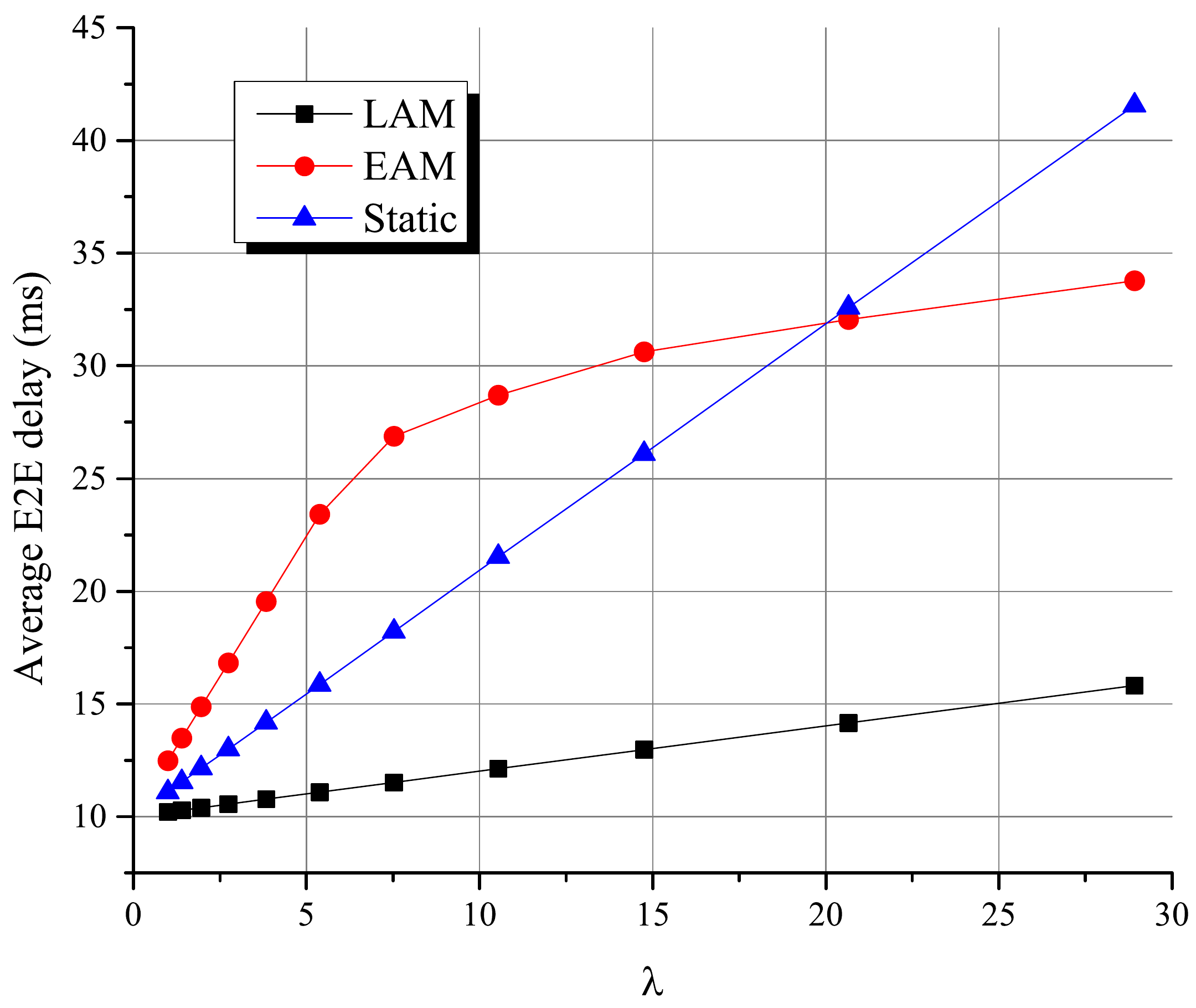}
  \caption{Average E2E delay by varying the value of $\lambda$.}
  \label{fig:perfm_lambda_delay}
\end{subfigure}
\hspace{0.2cm}
\begin{subfigure}{.4\textwidth}
  \centering
  \includegraphics[width=1.0\linewidth]{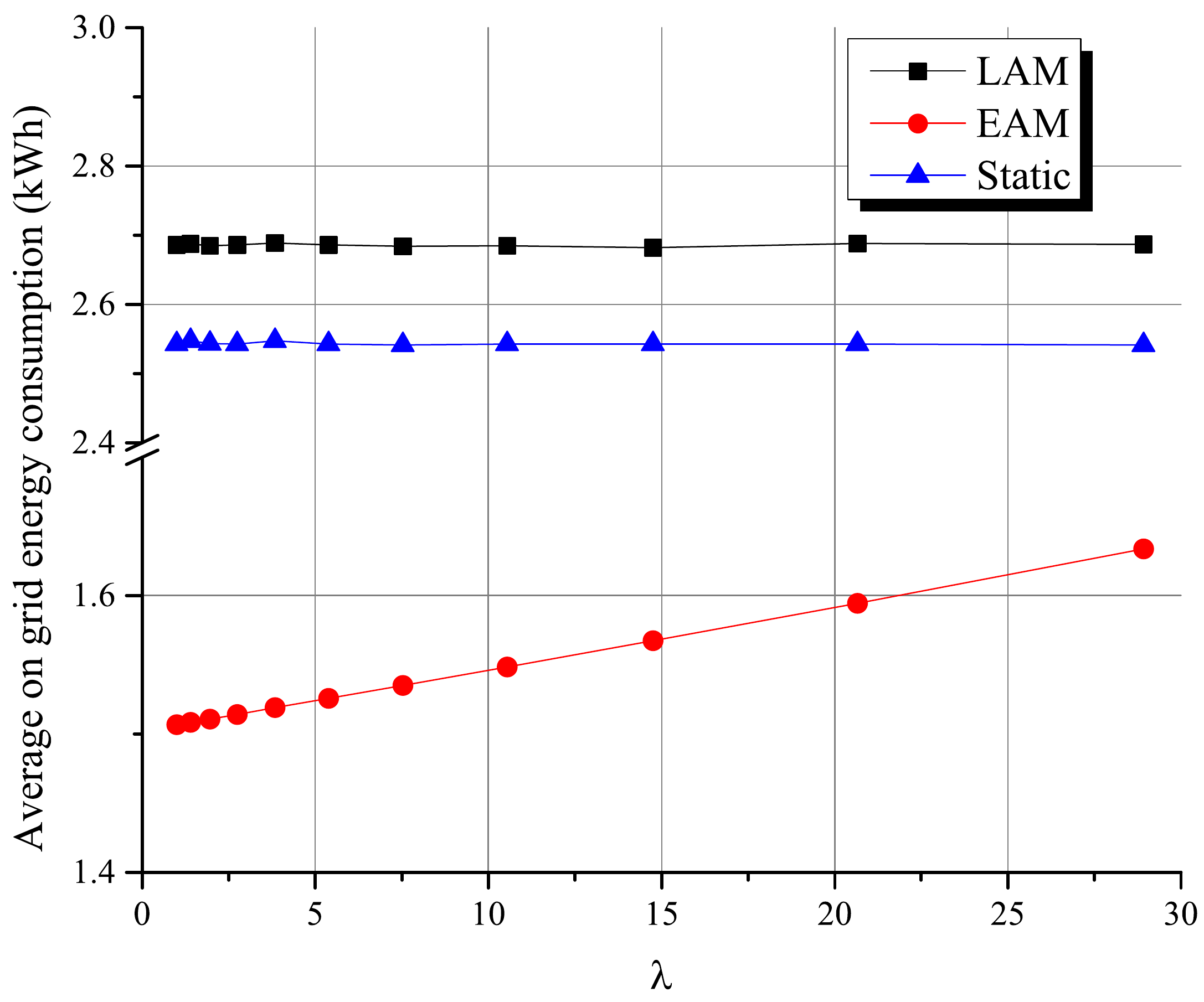}
  \caption{Average on-grid energy consumption by varying $\lambda$.}
  \label{fig:perfm_lambda_consumption}
\end{subfigure}
\caption{Simulation results.}
\label{fig:sim_result}
\end{figure*}

Fig. \ref{fig:overall_perfm} shows the average E2E delay between mobile devices and their proxy VMs, and the average on-grid energy consumption of the network during the monitoring period (i.e., six hours). Obviously, LAM incurs the lowest average E2E delay as compared to EAM and Static because if LAM is applied, once a mobile device roams from one BS into a new BS, its proxy VM may also be migrated into the cloudlet, which has the lowest E2E delay respect to the new BS. Yet, LAM incurs the highest on-grid energy consumption because many mobile devices would move to the same BS' coverage area, and thus their proxy VMs would be migrated to the cloudlet (that is connected to the BS). Consequently, the energy demand of this cloudlet would be significantly increased, and thus the cloudlet's green energy is subsequently drained. This triggers the cloudlet to pull energy from the grid. On the other hand, since energy demands are moving to the cloudlet, other cloudlets may have superfluous green energy, i.e., small value of energy gap. This unbalanced energy gap among different cloudlets increases the total on-grid energy consumption by applying LAM. As a comparison, EAM can balance the energy gap among cloudlets by migrating proxy VMs from the cloudlets with positive energy gap into the cloudlets with negative energy gap. Thus, as shown in Fig. \ref{fig:overall_perfm}, although EAM incurs higher average E2E delay than LAM, it saves 39.17\% and 35.74\% of on-grid energy consumption as compared to LAM and Static, respectively. Therefore, we conclude that there is a tradeoff between minimizing the average E2E delay and minimizing the on-grid energy consumption. Note that EAM can also guarantee the E2E delay between each proxy VM and its mobile devices to be less than the predefined threshold (i.e., the value of $\gamma$), as demonstrated in Fig. \ref{fig:overall_perfm_SLA},  where the average E2E delay violation rate\footnote{${average\ E2E\ delay\ violation\ rate} = \frac{N}{{\left|\bm{\mathcal{K}} \right|}}$, where $N$ refers to the number of the mobile devices with E2E delay larger than $\gamma$, and $\left|\bm{\mathcal{K}} \right|$ is the total number of mobile devices in the network.} for EAM during the monitoring period is 0\%. Yet, the average E2E delay violation rate of LAM and Static are 6.3\% and 33.6\%, respectively. Fig. \ref{fig:overall_perfm_SLA} also shows the maximum E2E delay among all the mobile devices and their proxy VMs during the monitoring period. Interestingly, although LAM incurs the lowest average E2E delay, the maximum E2E delay of LAM is larger than that of EAM. 

We further investigate how the amount of green energy generation (i.e., the value of $g_k$) affects the performance of the three methods. As shown in Fig. \ref{fig:perfm_green_consumption}, the average on-grid energy consumption by applying LAM and Static almost linearly decreases as the amount of green energy generation increases. Yet, the decrement of the average on-grid energy consumption by applying EAM is much faster than LAM and Static. This demonstrates that EAM can better utilize green energy by further balancing the energy gaps among cloudlets when more green energy is available. On the other hand, as shown in Fig. \ref{fig:perfm_green_delay}, the average E2E delay incurred by EAM is slightly increasing as the amount of green energy generation increases. This is because more available green energy in each cloudlet may result in more proxy VMs being migrated to the cloudlets (with negative gap), thus incurring long E2E delay. Note that the average E2E delay incurred by LAM and Static does not change as the amount of green energy generation increases.

In addition, we investigate how the amount of traffic load in the mobile core network affects the performance of the three methods. Note that increasing the traffic load in the mobile core network leads to increasing the E2E delay between a cloudlet and a BS. Thus, we use the value of $\lambda$ to reflect the traffic loads of the mobile core network, i.e., larger value of $\lambda$ implies heavier traffic load of the mobile core network, and vice versa. As shown in Fig. \ref{fig:perfm_lambda_delay}, when the value of $\lambda$ is small, the average E2E delay incurred by the three methods is similar. However, as the value of $\lambda$ becomes larger, the average E2E delay gap between LAM and EAM/Static becomes larger. Fig.\ref{fig:perfm_lambda_consumption} shows the average on grid energy consumption by varying the value of $\lambda$. The energy consumption of LAM and Static does not change since the value of $\lambda$ does not affect their migration strategies. Yet, the average on-grid energy consumption incurred by EAM is increasing as the value of $\lambda$ increases because a larger value of $\lambda$ causes less flexibility to balance the energy gap among cloudlets while satisfying Constraint \eqref{ct_4}.

\section{Future works}
Migrating proxy VMs among cloudlets can potentially reduce the E2E delay between mobile devices and their proxy VMs as well as the on-grid energy consumption of the whole network. However, the following issues need to be considered in order to design a more efficient proxy VM migration strategy:
\begin{itemize}[leftmargin=*]
\item{Proxy VM decomposition: A proxy VM is associated with a number of static/mobile IoT devices owned by the same user. Migrating the whole proxy VM when its mobile devices roam away can reduce the E2E delay between mobile devices and its proxy VM, but may increase the E2E delay between static devices and its proxy VM. Thus, before conducting migration, it is beneficial to decompose the proxy VM into two proxy VMs: one proxy VM continues to serve the static IoT devices, and the other proxy VM migrates among cloudlets as mobile IoT devices roams away \cite{Sun:2016:EdgeIoT}.}
\item{Migration overheads: Migrating a proxy VM among cloudlets introduces extra overheads. From the networking perspective, the migration overheads incur additional proxy VM migration traffic, which is determined by the size of the proxy VM, the provisioned bandwidth, etc. \cite{Sun:2016:PRIMAL}. From the energy consumption perspective, the migration overheads refer to the amount of energy incurred by the proxy VM migration. Migrating a proxy VM from the source cloudlet to the destination cloudlet may introduce non-negligible energy consumption from both source and destination cloudlets \cite{Strunk:2013:DLM}; thus, designing a proxy VM migration strategy without considering the migration energy consumption may significantly increase the total on-grid energy consumption \cite{Fan:2017:EDA}. From the performance of the proxy VM perspective, the migration overheads indicate the performance degradation of the proxy VM \cite{Anand:2013:VMP}. Specifically, conducting data sharing and analytics in a proxy VM consumes CPU, memory, and network resource of the proxy VM; meanwhile, proxy VM migration is considered as an expensive application, which consumes a significant amount of resources in the proxy VM. Thus, proxy VM migration can decelerate the speed of data sharing and analytics, which are conducted during the migration proxy.}
\item{Rightsizing cloudlets: The capacities of different cloudlets (i.e., the values of $\phi_k$) are assumed to be the same in the simulations. However, the capacity may vary among the cloudlets to further improve the performance (i.e., the average E2E delay) of the proxy VM migration strategies. For instance, the cloudlets located in the dense areas (such as train stations) may have higher capacities (to host more proxy VMs) than the cloudlets in sparse areas. In addition, the amount of green energy generation (i.e., the value of $g_k$) may also vary among the cloudlets to reduce the number of proxy VM migrations, i.e., the cloudlets with higher capacities could have more green energy generator units (e.g., larger sized solar panels to produce more green energy) than the cloudlets with lower capacities. Thus, it is beneficial to rightsizing cloudlets by optimizing the capacity and green energy generation for each cloudlet.}
\end{itemize} 
In the future, we will design an efficient proxy VM migration strategy by considering both the proxy VM decomposition and migration overheads to maximize the profit of IoT users/network providers. Also, we will implement the proposed semantic social IoT and proxy VM migration in our MEC lab. 

\section{Conclusion}

In this paper, we have introduced the MEIoT architecture to facilitate the IoT data sharing/analysis. We have illustrated four challenges in the current IoT system and proposed the corresponding solutions in the context of MEIoT. We have evaluated the performance of the two proposed proxy VM migration methods, i.e., LAM and EAM, via simulations. In addition, we have elicited future research directions of MEIoT. 

%\bibliographystyle{ieicetr}% bib style
%\bibliography{}% your bib database

\end{document}